\documentclass[sigconf]{acmart}
\AtBeginDocument{%
  \providecommand\BibTeX{{%
    \normalfont B\kern-0.5em{\scshape i\kern-0.25em b}\kern-0.8em\TeX}}}
\usepackage{xspace}
\usepackage{amsthm}
\usepackage{amsmath}
\usepackage{enumitem}
\usepackage{booktabs}
\usepackage{multirow}
\usepackage{graphicx}
\usepackage{float}
\usepackage{caption}
\usepackage{subcaption}
\usepackage{balance}
\usepackage[toc,page]{appendix}
\newcommand{\modelname}{\textsf{STOSA}\xspace}

\setcopyright{acmcopyright}
\copyrightyear{2022}
\acmYear{2022}
\setcopyright{acmcopyright}\acmConference[WWW '22]{Proceedings of the ACM Web
Conference 2022}{April 25--29, 2022}{Virtual Event, Lyon, France}
\acmBooktitle{Proceedings of the ACM Web Conference 2022 (WWW '22), April
25--29, 2022, Virtual Event, Lyon, France}
\acmPrice{15.00}
\acmDOI{10.1145/3485447.3512077}
\acmISBN{978-1-4503-9096-5/22/04}

\begin{document}

\title{Sequential Recommendation via Stochastic Self-Attention}

\author{Ziwei Fan*, Zhiwei Liu}\thanks{*Part of this work is done in Spotify Research}
\affiliation{%
  \institution{Department of Computer Science, University of Illinois at Chicago}
  \country{USA}
}
\email{{zfan20, zliu213}@uic.edu}

\author{Alice Wang, Zahra Nazari}
\affiliation{%
  \institution{Spotify}
  \country{USA}
}
\email{{alicew, zahran}@spotify.com}

\author{Lei Zheng}
\affiliation{%
  \institution{Pinterest Inc}
  \country{USA}
}
\email{lzheng@pinterest.com}

\author{Hao Peng}
\affiliation{%
  \institution{School of Cyber Science and Technology, Beihang University}
  \country{China}
}
\email{penghao@act.buaa.edu.cn}

\author{Philip S. Yu}
\affiliation{%
  \institution{Department of Computer Science, University of Illinois at Chicago}
  \country{USA}
}
\email{psyu@uic.edu}

\renewcommand{\shortauthors}{Ziwei Fan, et al.}

\begin{abstract}
Sequential recommendation models the dynamics of a user's previous behaviors in order to forecast the next item, and has drawn a lot of attention. Transformer-based approaches, 
which embed items as vectors and use dot-product self-attention to measure the relationship between items, demonstrate superior capabilities among existing sequential methods.
However, users' real-world sequential behaviors are \textit{\textbf{uncertain}} rather than deterministic, posing a significant challenge to present techniques. We further suggest that dot-product-based approaches cannot fully capture \textit{\textbf{collaborative transitivity}}, which can be derived in item-item transitions inside sequences and is beneficial for cold start items.
We further argue that BPR loss has no constraint on positive  and sampled negative items, which misleads the optimization.

We propose a novel \textbf{STO}chastic \textbf{S}elf-\textbf{A}ttention~(\modelname) to overcome these issues. STOSA, in particular, embeds each item as a stochastic Gaussian distribution, the covariance of which encodes the uncertainty. We devise a novel Wasserstein Self-Attention module to characterize item-item position-wise relationships in sequences, which effectively incorporates uncertainty into model training.
Wasserstein attentions also enlighten the collaborative transitivity learning as it satisfies triangle inequality. 
Moreover, we introduce a novel regularization term to the ranking loss, which assures the dissimilarity between positive and the negative items. Extensive experiments on five real-world benchmark datasets demonstrate the superiority of the proposed model over state-of-the-art baselines, especially on cold start items. 
The code is available in \url{https://github.com/zfan20/STOSA}.


\end{abstract}




\maketitle

\section{Introduction}
Recommender systems~\cite{liu2020basket, 10.1145/3459637.3482242, 10.1145/3459637.3482092, liu2021federated, gong2020attentional, wang2021pre, nazari2020recommending, li2020podcasts} become crucial components in web applications~\cite{lin2020fill,zhou2021intrinsic,zhou2021pure}, 
which provide personalized item lists by modeling interactions between users and items.
Sequential recommendation~(SR) attracts a lot of attention from both the academic community and industry due to its success and scalability. SR methods format each user's historical interactions as a sequence by sorting interactions chronologically. 
The goal of SR is to characterize users' evolving interests and predict the next preferred item. 

SR encodes users' dynamic interests by modeling item-item transition relationships in sequences.
Recent advancements in Transformer~\cite{vaswani2017attention, 10.1145/3404835.3463036} introduce the self-attention mechanism to reveal the position-wise item-item relationships, which leads to the state-of-the-art performance in SR.
SASRec is the pioneering work in proposing Transformer for sequential recommendation, which applies scaled dot-product self-attention to learn item-item correlation weights. 
BERT4Rec~\cite{sun2019bert4rec} adopts bi-directional modeling in sequences. 
TiSASRec~\cite{li2020time} and SSE-PT~\cite{wu2020sse} extend SASRec with additional time interval information and user regularization, respectively.

Despite the success of self-attention in sequential recommendation, we argue that methods based on dot-product self-attention
fail to incorporate: 1) \textit{dynamic uncertainty} and 2) \textit{collaborative transitivity}. 

Firstly, existing SR methods assume that dynamic user interests are deterministic. As such, the inferred user embeddings are fixed vectors in the latent space, which are insufficient to represent multifarious user interests, especially in the real-world dynamic environment.
Item transitions, which reflect the evolving process of user sequential behaviors, are sometimes hard to understand, and the two items in one item transition may not even lie in the same product category.
Hence, if a user has a significant portion of unexpected item transitions, modeling this user with a deterministic process achieves sub-optimal recommendations.
For example, in books recommendation, a user interested in science-fiction, romance, and biography is more uncertain than another user interested in thriller, horror, and fantasy.
Moreover, even two users share the same interest topics, the user with more fluctuated interests~(\textit{e.g.,} items in item transitions are in different topics) is more uncertain. 
Intuitively, users with greater interest dynamic variability are more uncertain. 
Therefore, dynamic uncertainty is a crucial component when we model user interests in a sequential environment.

Another limitation of the existing self-attention mechanism is that it fails to incorporate \textit{collaborative transitivity} in sequences. Collaborative transitivity can realize the latent similarity between items appearing in the same item-item transition pair but also inductively introduces additional collaborative similarities beyond limited item-item transitions in datasets. Thus, the collaborative transitivity can further alleviate cold-start item issues with the help of extra inducted collaborative similar items.
For example, given item transition pairs $(i_x \rightarrow i_y)$ and $(i_y \rightarrow i_z)$, 
we can conclude that $i_x$ and $i_y$ are close to each other and so as for $i_y$ and $i_z$. 
According to collaborative transitivity, $i_x$ and $i_z$ should also be close.
However, existing dot-product self-attention is unable to realize this collaborative closeness.
For example, given embeddings of $i_x=[0, 2]$, $i_y=[1, 1]$, $i_z=[2, 0]$, the dot-products of $(i_x, i_y)$ and $(i_y, i_z)$ are both 2, however, the dot-product between $i_x$ and $i_z$ is 0 because $i_x$ and $i_z$ transition pair is not observed. This issue becomes worse for unpopular items~(\textit{item cold start problem}) as the insufficient data for cold start items limits the set of collaborative neighbors. 

However, it is rather challenging to resolve the dynamic uncertainty and incorporate the collaborative transitivity into a SR model. 
Firstly, characterizing dynamic uncertainty among item transition relationships is still under-explored. Most existing self-attention SR models, \textit{e.g.,} SASRec~\cite{kang2018self} and BERT4Rec~\cite{sun2019bert4rec}, represent items as fixed vector embeddings, ignoring the uncertainty in sequential correlations. 
A recent work DT4SR~\cite{fan2021modeling} represents items as distributions, which proposes the mean and covariance embedding to model uncertainty in items. However, DT4SR is incapable of modeling dynamic uncertainty as it models item transition relationships via dot-product attention, which cannot incorporate such dynamic uncertainty. 

On top of modeling dynamic uncertainty, inducing collaborative transitivity remains a challenge. Existing works~\cite{kang2018self, sun2019bert4rec, rendle2012bpr, he2020lightgcn, tang2018personalized} based on the dot-product fail to satisfy the triangle inequality and consequently cannot accomplish the collaborative transitivity. Different from dot product, distances typically satisfy triangle inequality\footnote{$d(i_x, i_z) \leq d(i_x, i_y) + d(i_y, i_z)$}, which transits additional collaborative closeness and benefits a lot in item cold start issue. This assumption is theoretically supported in TransRec~\cite{he2017translation} and will also be empirically demonstrated in the following experimental results section.  Although some metric learning frameworks~\cite{li2020symmetric, zheng2019deep, he2017translation} propose distance functions to guarantee the triangle inequality, none of them can model dynamic uncertainty as well as collaborative transitivity in the sequential setting. The choice of distance function is pivotal to collaborative transitivity modeling.

Moreover, we argue that collaborative transitivity optimized by standard Bayesian Personalized Ranking~(BPR)\footnote{We only discuss BPR loss in this paper because it is the most widely used ranking loss for the top-N recommendation, but the same issue also happens to other losses, such as Hinge Loss.}~\cite{rendle2012bpr} loss fails to guarantee the dissimilarity between positive and negative items in BPR. BPR measures the difference between a user's preference scores on the positive item and a randomly sampled negative item. However, there is no guarantee that the positive item is further away from the negative item in the latent space. 
The incorporation of the distance between positive items and negatively sampled items is reasonable and necessary.
Typically, negative items are sampled from items that the user never shows interest with, even if randomly sampled negative items are not necessarily hard negative items~\cite{ying2018graph} due to data bias~\cite{chen2020bias}. A proper way to sample negative items is an important topic in the recommendation, but it is beyond this paper's scope.

To this end, we propose a new framework \textbf{STO}chastic \textbf{S}elf-\textbf{A}ttention~(\modelname), which comprises of three modules: (1) stochastic embeddings, (2) Wasserstein self-attention, and (3) regularization term in BPR loss.
Specifically, we model items as Gaussian distributions with stochastic embeddings, consisting of mean~(for base interests) and covariance~(for the variability of interests) embeddings.
On top of stochastic embeddings, we propose to use distances to measure item transitions, which is originated from metric learning~\cite{hsieh2017collaborative, Pitis2020An}. We propose a novel Wasserstein Self-attention layer, which measures attentions as scaled Wasserstein distances between items. 
We also introduce a novel regularization term in BPR loss to consider the distance between positive items and negatively sampled items. 
The contributions of this work are as follows: 
\begin{itemize}[leftmargin=*]
    \item To the best of our knowledge, \modelname is the first work proposing a Wasserstein Self-Attention to consider collaborative transitivity in SR. 
    \item We introduce stochastic embeddings to measure both base interests and the variability of interests inherent in user behaviors and improve BPR loss for SR with an additional regularization for constraining the distance between positive items and negatively sampled items.
    \item \modelname outperforms the state-of-the-art recommendation methods. The experimental results also demonstrate the effectiveness of \modelname on cold start items.
    \item Several visualizations verify the effectiveness of Wasserstein self-attention over the traditional scaled dot-product self-attention and justify the improvements for cold start items  by collaborative transitivity. 
\end{itemize}

\section{Related Work}
Several topics are closely related to our research problem. 
We first introduce some relevant works in the sequential recommendation, which is the primary problem setting in this paper.
As we use distance rather than dot-product as the metric, recommendation methods with metric learning will also be discussed. Finally, we will introduce some relevant works about using distributions as representations.
\subsection{Sequential Recommendation}
Sequential Recommendation~(SR) recommends the next item based on the chronological sequence of the user's historical interactions. The fundamental idea of SR is learning sequential patterns within consecutive interacted items. One representative work of SR is FPMC~\cite{rendle2010factorizing}. With the inspiration of Markov Chain's capability of learning item-item transitions, FPMC fuses the idea of Markov Chains with matrix factorization. FPMC learns the first-order item transition matrix, assuming that the next-item prediction is only relevant to the previous one item. Fossil~\cite{he2016fusing} extends this idea and considers higher order of item transitions. Another line of SR uses convolutional neural networks for sequence modeling, such as Caser~\cite{tang2018personalized}. Caser regards the embedding matrix of items in the sequence as an image and applies convolution operators with the motivation of capturing local item-item transitions.

The advancements of sequence modeling developed in deep neural networks inspire the adoption of Recurrent Neural Network~(RNN)~\cite{hochreiter1997long, quadrana2017personalizing, ma2019hierarchical, yan2019cosrec, zheng2019gated, peng2021ham} and Self-Attention mechanisms into SR~\cite{kang2018self, sun2019bert4rec, li2020time}. For example, GRU4Rec~\cite{hidasi2015session} proposes to use Gated Recurrent Units in the session-based recommendation. The success of self-attention-based Transformer~\cite{vaswani2017attention,zhang2021pdaln} and BERT~\cite{devlin2018bert,zhang2020mzet} inspires the community to investigate the possibility of self-attention in the sequential recommendation. Unlike Markov chain and RNN methods, self-attention utilizes attention scores from all item-item pairs in the sequence. SASRec~\cite{kang2018self} and BERT4Rec~\cite{sun2019bert4rec} both demonstrate the effectiveness of self-attention with the state-of-the-art performance in next-item recommendation.

\subsection{Metric Learning for Recommendation}
Metric learning explores a proper distance function to measure the dissimilarity between objects, such as Euclidean distance, Mahalanobis distance~\cite{mclachlan1999mahalanobis} and Graph distance~\cite{gao2010survey}. A crucial property that differentiates distance and dot-product as metrics is that distances usually satisfy the triangle inequality. Triangle inequality, as an inductive bias~\cite{Pitis2020An} for distances, is useful when data sparsity issue exists~\cite{he2017translation}. One early work on metric learning for recommendation is CML~\cite{hsieh2017collaborative}. CML proposes a hinge loss on minimizing the L2 distance between embeddings of the user and interacted items. LRML~\cite{tay2018latent} then demonstrates the geometric restriction of CML and introduces a latent relation as a translation vector in the distance calculation. TransRec~\cite{he2017translation} borrows the idea of knowledge embedding and also develops a translation vector for the sequential recommendation. SML~\cite{li2020symmetric} is the state-of-the-art metric learning recommendation method. It introduces an additional item-centric metric and the adaptive margin on top of CML.

\subsection{Distribution Representations}
Representing objects~(e.g., words, nodes, and items) as distributions has been attracting interest from the research community~\cite{vilnis2014word, bojchevski2018deep, sun2018gaussian, he2015learning}. Distribution representations introduce uncertainties and provide more flexibility compared with one single fixed embedding. DVNE~\cite{zhu2018deep} utilizes Gaussian distribution as the node embedding in graphs and proposes a deep variational model for higher-order proximity information propagation. TIGER~\cite{qian2021conceptualized} and \cite{vilnis2014word} represent words as Gaussians, and TIGER also introduces Gaussian attention for better learning the entailment relationship among words. PMLAM~\cite{ma2020probabilistic} and DDN~\cite{zheng2019deep} both propose to use Gaussian distributions to represent users and items. DDN learns the mean and covariance embeddings with two neural networks. DT4SR~\cite{fan2021modeling} is the most relevant work, which represents items as distributions and learns mean and covariance with separate Transformers.

\section{Preliminaries and Discussions}
In this section, we first formulate the SR problem and then introduce the self-attention mechanism for solving this problem. 
\subsection{Problem Definition}
Given a set of users $\mathcal{U}$ and items $\mathcal{V}$, and their associated interactions, 
we can sort the interacted items of each user $u\in\mathcal{U}$ chronologically in a sequence
as $\mathcal{S}^{u}={[v^{u}_1,v^u_2,\dots,v^u_{|\mathcal{S}^{u}|}]}$, where $v^{u}_i\in\mathcal{V}$ denotes the $i$-th interacted item in the sequence. 
The goal of SR is to recommend a top-N ranking list of items 
as the potential
next items in a sequence.
Formally, we should predict $p\left( v_{|\mathcal{S}^{u}|+1}^{(u)}=v \left|  \mathcal{S}^{u} \right.\right)$.

\subsection{Self-Attention for Recommendation}


Since we adopt the self-attention mechanism as the backbone of sequence encoder, 
we introduce it before proposing our model.
The intuition of the self-attention mechanism is that items in sequences are correlated but of distinct importance to the items at different positions in a sequence. 
Specifically,
given a user's action sequence $\mathcal{S}^u$ and the maximum sequence length $n$, the sequence is first truncated by removing earliest items if $|\mathcal{S}^u|>n$ or padded with 0s to get a fixed length sequence $s=(s_1, s_2, \dots, s_n)$. An item embedding matrix $\mathbf{M}\in \mathbb{R}^{|\mathcal{V}|\times d}$ is defined, where $d$ is the number of latent dimensions. 
A trainable
positional embedding $\mathbf{P}\in \mathbb{R}^{n\times d}$ is added to sequence embedding matrix as:
\begin{equation}
    \label{eq:seq_emb}
    \hat{\mathbf{E}}_{\mathcal{S}^{u}} = [\mathbf{m}_{s_1}+\mathbf{p}_{s_1}, \mathbf{m}_{s_2}+\mathbf{p}_{s_2}, \dots, \mathbf{m}_{s_n}+\mathbf{p}_{s_n}].
\end{equation}
Specifically, self-attention uses dot-products between items in the sequence to infer their correlations, which are as follows:
\begin{equation}
\label{eq:sa}
    \text{SA}(\mathbf{Q}, \mathbf{K}, \mathbf{V}) = \text{softmax}\left(\frac{\mathbf{Q}\mathbf{K}^\top}{\sqrt{d}}\right) \mathbf{V}, 
\end{equation}
where $\mathbf{Q}=\hat{\mathbf{E}}_{\mathcal{S}^{u}}\mathbf{W}^Q$, $\mathbf{K}=\hat{\mathbf{E}}_{\mathcal{S}^{u}}\mathbf{W}^K$, and $\mathbf{V}=\hat{\mathbf{E}}_{\mathcal{S}^{u}}\mathbf{W}^V$. As both $\mathbf{Q}$ and $\mathbf{K}$ use the same input sequence, the scaled dot-product component can learn the latent correlation between items. Additionally, other components in Transformer are utilized in SASRec, including the point-wise feed-forward network, residual connection, and layer normalization.


\begin{figure*}
         \centering
         \includegraphics[width=0.77\textwidth]{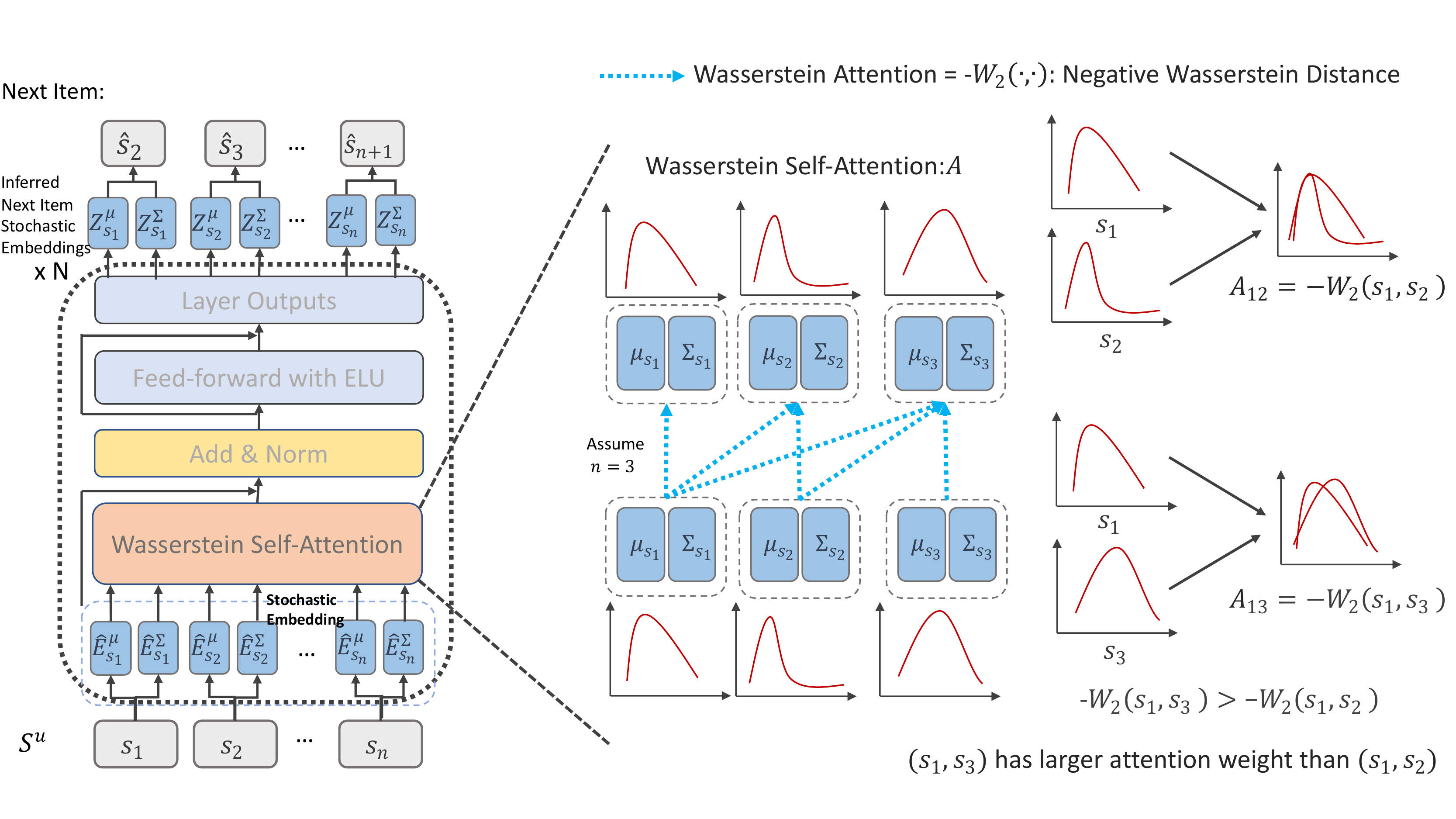}
         \caption{Model Architecture of the proposed \modelname. $s_i$ denotes the item in the position $i$ and $\hat{s}_{i+1}$ indicates the output inferred next item in $(i+1)$-th position. We propose stochastic embeddings to consider dynamic uncertainty information and introduce a novel Wasserstein Self-Attention layer for capturing collaborative transitivity signals. We introduce Feed-forward networks with ELU activation and guarantee the positive definite property of covariances. }
         \label{fig:model_architecture}
    
\end{figure*}

\section{Proposed Model}
In this section, we introduce stochastic self-attention (\modelname) to overcome limitations of existing dot-product self-attention, as shown in Figure~\ref{fig:model_architecture}. We first represent items as stochastic embeddings with Elliptical Gaussian distributions, 
comprised of the mean embedding and covariance embedding. Then we develop a novel Wasserstein self-attention module based on the Wasserstein distance to infer the stochastic sequence embeddings. A Wasserstein distance is adopted to measure the dissimilarity between items in the sequence with uncertainty signals.
Finally, we incorporate a novel regularization term measuring the distance between positive and negative items into the standard BPR loss.

\subsection{Stochastic Embedding Layers}
We introduce uncertainty into item embeddings by representing items as distributions.
Differing deterministic vector representation, modeling items as stochastic distributions covers larger space for including more collaborative neighbors.
Specifically, we use multi-dimensional elliptical Gaussian distributions to represent items.
An elliptical Gaussian distribution is governed by a mean vector and a covariance vector\footnote{The covariance of elliptical Gaussian distribution is a diagonal matrix, therefore the diagonal values can be viewed as a vector.}, 
where covariance introduces the potential uncertainty of the item.
For all items, we define a mean embedding table $\mathbf{M}^{\mu}\in \mathbb{R}^{|\mathcal{V}|\times d}$ and the covariance embedding table $\mathbf{M}^{\Sigma}\in \mathbb{R}^{|\mathcal{V}|\times d}$. As mean and covariance identify different signals, we thus introduce separate positional embeddings for mean and covariance $\mathbf{P}^{\mu}\in \mathbb{R}^{n\times d}$ and $\mathbf{P}^{\Sigma}\in \mathbb{R}^{n\times d}$, respectively. In analogy to Eq.~(\ref{eq:seq_emb}), we can obtain mean and covariance sequence embeddings of user $u$ as:
\begin{equation}
\begin{aligned}
    \label{eq:sequence_embed}
    \hat{\mathbf{E}}^{\mu}_{\mathcal{S}^{u}} &=
    [\hat{\mathbf{E}}^{\mu}_{s_1}, \hat{\mathbf{E}}^{\mu}_{s_2}, \dots, \hat{\mathbf{E}}^{\mu}_{s_n}]
    =[\mathbf{m}^{\mu}_{s_1}+\mathbf{p}^{\mu}_{s_1}, \mathbf{m}^{\mu}_{s_2}+\mathbf{p}^{\mu}_{s_2}, \dots, \mathbf{m}^{\mu}_{s_n}+\mathbf{p}^{\mu}_{s_n}],\\
    \hat{\mathbf{E}}^{\Sigma}_{\mathcal{S}^{u}} &=
    [\hat{\mathbf{E}}^{\Sigma}_{s_1}, \hat{\mathbf{E}}^{\Sigma}_{s_2}, \dots, \hat{\mathbf{E}}^{\Sigma}_{s_n}]
    =[\mathbf{m}^{\Sigma}_{s_1}+\mathbf{p}^{\Sigma}_{s_1}, \mathbf{m}^{\Sigma}_{s_2}+\mathbf{p}^{\Sigma}_{s_2}, \dots, \mathbf{m}^{\Sigma}_{s_n}+\mathbf{p}^{\Sigma}_{s_n}].
\end{aligned}
\end{equation}
For example, for the first item $s_1$ in the sequence, its stochastic embedding is represented as a $d$-dimensional Gaussian distribution $\mathcal{N}(\mu_{s_1}, \Sigma_{s_1})$, where $\mu_{s_1}=\hat{\mathbf{E}}^{\mu}_{s_1}$ and $\Sigma_{s_1}=diag(\hat{\mathbf{E}}^{\Sigma}_{s_1}))\in\mathbb{R}^{d\times d}$.

\subsection{Wasserstein Self-Attention Layer}
There remain challenges in modeling sequential dynamics with stochastic embeddings. First, it remains problematic to model dynamics of item transitions with distributions while still satisfying the triangle inequality. Secondly, the aggregation of these sequential signals to obtain the sequence's representation~(i.e., the user's representation) is still not resolved. 
To tackle both challenges, we introduce Wasserstein distances as attention weights to measure the pair-wise relationships between items in the sequence, and we also adopt the linear combination property of Gaussian distributions~\cite{dwyer1958generalizations} to aggregate historical items and obtain the sequence representation.

\subsubsection{Wasserstein Attention}
We propose a novel variant of self-attention adaptive to stochastic embeddings. We first denote $\mathbf{A}\in\mathbb{R}^{n\times n}$ as the self-attention values. $A_{kt}$ denotes the attention value between item $s_k$ and item $s_t$ in $k$-th and $t$-th positions in the sequence, where $k\leq t$ with the consideration of causality, repspectively. According to Eq.~(\ref{eq:sa}), the attention weight of traditional self-attention is calculated as:
\begin{equation}
    \mathbf{A}_{kt} = \mathbf{Q}_{k}\mathbf{K}_{t}^\top/\sqrt{d}.
\end{equation}
However, dot-product is not designed for measuring the discrepancy between distributions~(i.e., stochastic embeddings) and fails to satisfy triangle inequality. Instead, we adopt Wasserstein distance\footnote{We also tried Kullback–Leibler~(KL) divergence, but it achieves worse performance and inferior inference efficiency as well as violates triangle inequality.}~\cite{ruschendorf1985wasserstein} to measure the distance between stochastic embeddings of two items. Formally, given two items $s_k$ and $s_t$, the corresponding stochastic embeddings are $\mathcal{N}(\mu_{s_k}, \Sigma_{s_k})$ and $\mathcal{N}(\mu_{s_t}, \Sigma_{s_t})$, where $\mu_{s_k} = \hat{\mathbf{E}}^{\mu}_{s_k}W^{\mu}_K$, $\Sigma_{s_k}=\text{ELU}\left(diag(\hat{\mathbf{E}}^{\Sigma}_{s_k}W^{\Sigma}_K)\right)+1$, $\mu_{s_t} = \hat{\mathbf{E}}^{\mu}_{s_t}W^{\mu}_Q$, $\Sigma_{s_t}=\text{ELU}\left(diag(\hat{\mathbf{E}}^{\Sigma}_{s_t}W^{\Sigma}_Q)\right)+1$. Exponential Linear Unit~(ELU) maps inputs into $[-1, +\infty)$. It is used to guarantee the positive definite property of covariance. We define the attention weight as the negative 2-Wasserstein distance $W_2(\cdot, \cdot)$ is measured as follows,
\begin{equation}
\begin{split}
\label{eq:wass_dist_att}
    \mathbf{A}_{kt} &= -(W_2(s_k, s_t))\\
    &= -\left(||\mu_{s_k}-\mu_{s_t}||^2_2 + \text{trace}\left(\Sigma_{s_k}+\Sigma_{s_t}-2(\Sigma_{s_k}^{1/2}\Sigma_{s_k}\Sigma_{s_t}^{1/2})^{1/2}\right)\right),
\end{split}
\end{equation}
\textbf{Why Wasserstein distance?} There are several advantages of using Wasserstein distance. First, Wasserstein distance measures the distance between distributions, with the capability of measuring the dissimilarity of items with uncertainty information. Secondly, Wasserstein distance satisfies triangle inequality~\cite{clement2008elementary} and can capture collaborative transitivity inductively in sequence modeling. Finally, Wasserstein distance also enjoys the advantage of a more stable training process as it provides a smoother measurement when two distributions are non-overlapping~\cite{kolouri2018sliced}, which in SR means two items are far away from each other. However, KL divergence will produce an infinity distance, causing numerical instability. 

Note that Eq.~(\ref{eq:wass_dist_att}) can be computed with batch matrix multiplications without sacrificing computation and space efficiency compared with traditional self-attention, which will be discussed in the complexity analysis section of Appendices.

\subsubsection{Wasserstein Attentive Aggregation}
The output embedding of the item in each position of the sequence is the weighted sum of embeddings from previous steps, where weights are normalized attention values $\tilde{\mathbf{A}}$ as:
\begin{equation}
    \tilde{\mathbf{A}}_{kt} = \frac{\mathbf{A}_{kt}}{\sum_{j=1}^{t}\mathbf{A}_{jt}}.
\end{equation}
As each item is represented as a stochastic embedding with both mean and covariance, the aggregations of mean and covariance are different. We adopt the linear combination property of Gaussian distribution~\cite{dwyer1958generalizations}, which is as follows,
\begin{equation}
\begin{split}
    \mathbf{z}^{\mu}_{s_t} = \sum_{k=1}^{t} \tilde{\mathbf{A}}_{kt} \mathbf{V}^{\mu}_k, \:\text{and}\:
    \mathbf{z}^{\Sigma}_{s_t} = \sum_{k=0}^{t} \tilde{\mathbf{A}}_{kt}^2 \mathbf{V}^{\Sigma}_k,
\end{split}
\end{equation}
where $\mathbf{V}^{\mu}_{s_k}=\hat{\mathbf{E}}^{\mu}_{s_k}W^{\mu}_V$,  $\mathbf{V}^\Sigma_{s_k}=diag(\hat{\mathbf{E}}^{\Sigma}_{s_k}))W^{\Sigma}_V$, and $k\leq t$ for causality. The outputs $\mathbf{Z}^{\mu}=(\mathbf{z}^{\mu}_{s_1}, \mathbf{z}^{\mu}_{s_2},\dots,\mathbf{z}^{\mu}_{s_n})$ and $\mathbf{Z}^{\Sigma}=(\mathbf{z}^{\Sigma}_{s_1}, \mathbf{z}^{\Sigma}_{s_2},\dots,\mathbf{z}^{\Sigma}_{s_n})$ together form the newly generated sequence's stochastic embeddings, which aggregates historical sequential signals with awareness of uncertainty.
\subsection{Feed-Forward Network and Layer Outputs}
The self-attention and the aggregation learn relationships in linear transformation. However, non-linearity can capture more complex relationships. We apply two point-wise fully connected layers with an ELU activation to introduce non-linearity in learning stochastic embeddings:
\begin{equation}
\begin{split}
    \text{FFN}^{\mu}(\mathbf{z}^{\mu}_{s_t}) &= \text{ELU}(\mathbf{z}^{\mu}_{s_t}W^{\mu}_1+b^{\mu}_1)W^{\mu}_2 + b^{\mu}_2,\\
    \text{FFN}^{\Sigma}(\mathbf{z}^{\Sigma}_{s_t}) &= \text{ELU}(\mathbf{z}^{\Sigma}_{s_t}W^{\Sigma}_1+b^{\Sigma}_1)W^{\Sigma}_2 + b^{\Sigma}_2,
\end{split}
\end{equation}
where $W^{*}_1\in\mathbb{R}^{d\times d}$, $W^{*}_2\in\mathbb{R}^{d\times d}$, $b^{*}_1\in\mathbb{R}^{d}$, and $b^{*}_2\in\mathbb{R}^{d}$ are learnable parameters and $*$ can be $\mu$ or $\Sigma$. We adopt ELU instead of ReLU because of the numerical stability of ELU. We also adopt other components like~\cite{kang2018self, sun2019bert4rec, vaswani2017attention}, such as residual connection, layer normalization, and dropout layers, the layer outputs are,
\begin{equation}
\begin{split}
    \mathbf{Z}^{\mu}_{s_t} &= \mathbf{z}^{\mu}_{s_t} + \text{Dropout}(\text{FFN}^{\mu}(\text{LayerNorm}(\mathbf{z}^{\mu}_{s_t}))),\\
    \mathbf{Z}^{\Sigma}_{s_t} &= \text{ELU}\left(\mathbf{z}^{\Sigma}_{s_t} + \text{Dropout}(\text{FFN}^{\Sigma}(\text{LayerNorm}(\mathbf{z}^{\Sigma}_{s_t})))\right)+1.
\end{split}
\end{equation}
We adopt  ELU activation and ones addition to covariance embeddings to guarantee the positive definite property of covariance. Note that if we stack more layers, $\mathbf{Z}^{\mu}$ and $\mathbf{Z}^{\Sigma}$ can be inputs of the next Wasserstein self-attention layer. We ignore the layer superscript for avoiding over-complex symbolization. 

\subsection{Prediction Layer}
We predict the next item based on output embeddings $\mathbf{Z}^{\mu}$ and $\mathbf{Z}^{\Sigma}$ from last layer if we stack several layers. We adopt the similar shared item embedding strategy in~\cite{kang2018self,sun2019bert4rec, li2020time} for reducing model size and the risk of overfitting. Formally, for the item $s_t$ in the $t$-th position of the sequence, the prediction score of next item $j$ at $(t+1)$-th position is formulated as 2-Wasserstein distance of two distributions $\mathcal{N}(\mu_{s_t}, \Sigma_{s_t})$ and $\mathcal{N}(\mu_{j}, \Sigma_{j})$,
\begin{equation}
\label{eq:pred}
    d_{s_t, j} = W_2(s_t, j),
\end{equation}
where $\mu_{s_t}=\mathbf{Z}^{\mu}_{s_t}$ and $\Sigma_{s_t} = \mathbf{Z}^{\Sigma}_{s_t}$ are inferred representations given $(s_1, s_2, \cdots, s_t), 1\leq t\leq n$; $\mu_j=\mathbf{M}^{\mu}_j$ and $\Sigma_{j}=\mathbf{M}^{\Sigma}_j$ are embeddings indexed from input stochastic embedding tables $\mathbf{M}^{\mu}$ and $\mathbf{M}^{\Sigma}$. 

For evaluation, different from dot-product methods, a smaller distance score indicates a higher probability of the next item. We thus generate the top-N recommendation list by sorting scores in ascending order.

\subsection{BPR Loss with Positive v.s. Negative}
We adopt the standard BPR loss~\cite{rendle2012bpr} as base loss for measuring the ranking prediction error. However, BPR loss fails to consider the distance between the positive item and the negative sampled item. Therefore, we introduce a regularization term to enhance such distances as follows:
\begin{equation}
    \ell_{pvn}(s_t, j^{+}, j^{-}) = \left[d_{s_t, j^{+}} - d_{j^{+}, j^{-}}\right]_{+},
\end{equation}
where $[x]_{+}=\text{max}(x, 0)$ is the standard hinge loss, $j^{+}$ is the ground truth next item, and $j^{-}$ is a randomly sampled negative item from items that the user never interacts with. The intuition behind $\ell_{pvn}(t, j^{+}, j^{-})$ is that the distance between positive item and negative item $d_{j^{+}, j^{-}}$ has to be larger than the prediction distance $d_{s_t, j^{+}}$. Otherwise, when $d_{j^{+}, j^{-}} < d_{s_t, j^{+}}$, it becomes counter-intuitive. The inequality $d_{j^{+}, j^{-}} < d_{s_t, j^{+}}$ indicates that the positive item $j^{+}$ is closer with negative item $j^{-}$ than with $s_t$ while $s_t$, as the previous item of $j^{+}$, should have a smaller distance with $j^{+}$ instead. We incorporate this hinge loss as a regularization term with BPR loss into the final loss as follows,
\begin{equation}
\label{eq:loss_obj}
    L = \sum_{\mathcal{S}^u\in\mathcal{S}}\sum_{t=1}^{|\mathcal{S}^u|}-\log (\sigma(d_{s_t, j^{-}} - d_{s_t, j^{+}})) + \lambda\ell_{pvn}(s_t, j^{+}, j^{-}) +\beta||\Theta||_2^2.
\end{equation}
We minimize $L$ and optimize all learnable parameters $\Theta$ with Adam optimizer~\cite{DBLP:journals/corr/KingmaB14}. In the ideal case, the second term $\lambda\ell(s_t, j^{+}, j^{-})$ becomes 0, which means $s_t$ is close to $j^{+}$ but both $s_t$ and $j^{+}$ are far away from $j^{-}$.


\section{Experiments}
In this section, we validate the effectiveness of the proposed \modelname in several aspects by presenting experimental results and comparisons. The experiments answer the following research questions~(RQs): 
\begin{itemize}[leftmargin=*]
    \item \textbf{RQ1}: Does \modelname provide better recommendations than baselines?
    \item \textbf{RQ2}: What is the influence of Wasserstein self-attention and the regularization term $\ell_{pvn}$?
    \item \textbf{RQ3}: Does \modelname help the item cold start issue?
    \item \textbf{RQ4}: What is the distinction between dot-product and Wasserstein attention?
    \item \textbf{RQ5}: Why \modelname can alleviate the item cold-start issue?
\end{itemize}

\subsection{Datasets}
We evaluate the proposed \modelname on five public benchmark datasets from Amazon review datasets across various domains, with more than 1.2 million users and 63k items in total. Amazon datasets are known for high sparsity and have several categories of rating reviews. We use timestamps of each rating to sort interactions of each user to form the sequence. The latest interaction is used for testing, and the last second one is used for validation. Following~\cite{sun2019bert4rec, kang2018self, fan2021modeling, he2017translation, li2020time, wu2020sse}, we also adopt the 5-core settings by filtering out users with less than 5 interactions. We treat the presence of ratings as positive interactions. Details of datasets statistics and preprocessing steps are in the Appendices.

\subsection{Evaluation}
For each user, we sort the prediction scores calculated by Eq.~(\ref{eq:pred}) in ascending order to generate the top-N recommendation list. We \textbf{rank all items} instead of the biased sampling evaluation~\cite{krichene2020sampled}.
We adopt the standard top-N ranking evaluation metrics, Recall@N, NDCG@N, and Mean Reciprocal Rank~(MRR). Recall@N measures the average number of positive items being retrieved in the generated top-N recommendation list for each user. NDCG@N extends Recall@N by also considering the positions of retrieved positive items in the top-N list. MRR measures the ranking performance in the entire ranking list instead of top-N. We report the averaged metrics over all users. We report the performances when $N=1$ and $N=5$. 
\subsection{Baselines}
We compare the proposed \modelname with the following baselines in three groups. The first group includes static recommendation methods, which ignore the sequential order, including BPR~\cite{rendle2012bpr} and LightGCN~\cite{he2020lightgcn}. The second group of baselines consist of recommendation methods based on metric learning, including CML~\cite{hsieh2017collaborative}, SML~\cite{li2020symmetric}. The third group includes sequential recommendation methods: TransRec~\cite{he2017translation}, Caser~\cite{tang2018personalized}, SASRec~\cite{kang2018self}, DT4SR~\cite{fan2021modeling}, and BERT4Rec~\cite{sun2019bert4rec}. 

For all baselines, we search the embedding dimension in $\{64, 128\}$. As the proposed model has both mean and covariance embeddings, we only search for $\{32, 64\}$ for \modelname for the fair comparison. More details of hyper-parameters grid search are in Appendices. 

\begin{table*}[]
\centering
\caption{Overall Performance Comparison Table. The best and second-best results are bold and underlined, respectively. `OOM' means the out-of-memory error. `Improve.' is the relative improvement against the second-best baseline performance. }
\label{tab:overal_perf}
\resizebox{\textwidth}{!}{%
\begin{tabular}{ccccccccccccc}
\hline
Dataset & Metric & BPRMF & LightGCN & CML & SML & TransRec & Caser & SASRec & DT4SR & BERT4Rec & \multicolumn{1}{l}{\modelname} & Improv. \\ \hline
\multirow{4}{*}{Home} & Recall@1 & 0.0029 & 0.0026 & 0.0025 & 0.0026 & 0.0018 & OOM & \underline{ 0.0046} & 0.0029 & 0.0029 & \textbf{0.0053} & +13.63\% \\
 & Recall@5 & 0.0096 & 0.0095 & 0.0076 & 0.0084 & 0.0063 & OOM & 0.0127 & \underline{0.0129} & 0.0105 & \textbf{0.0133} & +3.16\% \\
 & NDCG@5 & 0.0062 & 0.0060 & 0.0059 & 0.0056 & 0.0040 & OOM & \underline{0.0087} & 0.0082 & 0.0067 & \textbf{0.0093} & +6.76\% \\
 & MRR & 0.0073 & 0.0071 & 0.0062 & 0.0061 & 0.0052 & OOM & \underline{0.0094} & 0.0093 & 0.0092 & \textbf{0.0100} & +6.17\% \\ \hline
\multirow{4}{*}{Beauty} & Recall@1 & 0.0082 & 0.0064 & 0.0072 & 0.0069 & 0.0085 & 0.0112 & 0.0129 & \underline{0.0143} & 0.0119 & \textbf{0.0193} & +35.11\% \\
 & Recall@5 & 0.0300 & 0.0287 & 0.0249 & 0.0279 & 0.0321 & 0.0309 & 0.0416 & \underline{0.0449} & 0.0396 & \textbf{0.0504} & +12.15\% \\
 & NDCG@5 & 0.0189 & 0.0174 & 0.0184 & 0.0173 & 0.0204 & 0.0214 & 0.0274 & \underline{0.0296} & 0.0257 & \textbf{0.0351} & +18.45\% \\
 & MRR & 0.0216 & 0.0203 & 0.0198 & 0.0191 & 0.0236 & 0.0231 & 0.0291 & \underline{0.0323} & 0.0294 & \textbf{0.0360} & +11.54\% \\ \hline
\multirow{4}{*}{Tools} & Recall@1 & 0.0062 & 0.0071 & 0.0048 & 0.0055 & 0.0059 & 0.0056 & \underline{0.0103} & \underline{0.0103} & 0.0059 & \textbf{0.0120} & +15.81\% \\
 & Recall@5 & 0.0216 & 0.0231 & 0.0129 & 0.0156 & 0.0210 & 0.0129 & 0.0284 & \underline{0.0289} & 0.0189 & \textbf{0.0312} & +7.85\% \\
 & NDCG@5 & 0.0139 & 0.0152 & 0.0096 & 0.0107 & 0.0134 & 0.0091 & 0.0194 & \underline{0.0196} & 0.0123 & \textbf{0.0217} & +11.04\% \\
 & MRR & 0.0154 & 0.0170 & 0.0107 & 0.0118 & 0.0152 & 0.0106 & \underline{0.0207} & 0.0206 & 0.0160 & \textbf{0.0226} & +9.81\% \\ \hline
\multirow{4}{*}{Toys} & Recall@1 & 0.0084 & 0.0077 & 0.0072 & 0.0102 & 0.0062 & 0.0089 & 0.0193 & \underline{0.0202} & 0.0110 & \textbf{0.0240} & +18.88\% \\
 & Recall@5 & 0.0301 & 0.0266 & 0.0249 & 0.0283 & 0.0222 & 0.0240 & \underline{0.0551} & 0.0550 & 0.0300 & \textbf{0.0577} & +4.86\% \\
 & NDCG@5 & 0.0194 & 0.0173 & 0.0154 & 0.0195 & 0.0143 & 0.0210 & \underline{0.0377} & 0.0360 & 0.0206 & \textbf{0.0412} & +9.45\% \\
 & MRR & 0.0216 & 0.0200 & 0.0178 & 0.0210 & 0.0166 & 0.0221 & 0.0385 & \underline{0.0387} & 0.0244 & \textbf{0.0415} & +7.35\% \\ \hline
\multirow{4}{*}{Office} & Recall@1 & 0.0073 & 0.0088 & 0.0096 & 0.0090 & 0.0100 & 0.0069 & 0.0198 & \underline{0.0206} & 0.0137 & \textbf{0.0234} & +13.59\% \\
 & Recall@5 & 0.0214 & 0.0226 & 0.0249 & 0.0190 & 0.0343 & 0.0302 & \underline{0.0656} & 0.0630 & 0.0485 & \textbf{0.0677} & +3.20\% \\
 & NDCG@5 & 0.0144 & 0.0157 & 0.0172 & 0.0140 & 0.0219 & 0.0186 & \underline{0.0428} & 0.0421 & 0.0309 & \textbf{0.0461} & +7.71\% \\
 & MRR & 0.0162 & 0.0181 & 0.0191 & 0.0164 & 0.0263 & 0.0268 & 0.0457 & \underline{0.0475} & 0.0408 & \textbf{0.0502} & +5.68\% \\ \hline
\end{tabular}%
}
\end{table*}

\subsection{Overall Comparison~(RQ1 and RQ2)}
We compare the performance of all models in Table~\ref{tab:overal_perf} and demonstrate the effectiveness of \modelname. We interpret the results with following observations:
\begin{itemize}[leftmargin=*]
    \item \modelname obtains the best performance against all baselines in all metrics, especially in top-1 recommendation~(Recall@1). The relative improvements range from 3.16\% to 35.11\% in all metrics, demonstrating the superiority of \modelname. We can also observe that improvements are consistent in MRR for measuring the entire recommendation list, ranging from 5.68\% to 11.54\%. We attribute improvements to several factors of \modelname: (1). distribution representations help expand the latent interaction space of items to better understand uncertainty and flexibility; (2). collaborative transitivity enhances the discovery and induction of collaborative signals inherent in item-item transitions; (3). the newly introduced $\ell_{pvn}$ loss poses an additional constraint, which restrains distances between positive items and negative items to be no larger than ones of positive item transitions.
    \item Static methods, including BPRMF, LightGCN, CML, and SML, perform worse than sequential methods. This phenomenon verifies the necessity of temporal order information for the recommendation. Among all static methods, BPRMF, LightGCN achieve the best performances in different datasets. BPRMF and LightGCN perform better in Tools and Office datasets while achieving comparative results in other datasets. The reason for inferior performances of metric learning methods~(CML and SML) might be the norm constraint of embeddings as both models will normalize all embeddings to have the norm of one. 
    \item Among all sequential baselines, SASRec and DT4SR perform the best. DT4SR outperforms SASRec in three of five datasets, indicating the necessity of distribution representations and modeling uncertainty information in sequential recommendation. BERT4Rec fails to achieve satisfactory performance potentially due to the loss inconsistency between the adopted Cloze objective and the recommendation task. The comparison between Caser and Transformer-based methods demonstrates the effectiveness of self-attention in sequential modeling for recommendation. 
\end{itemize}

\begin{figure*}[ht]
     \centering
     \begin{subfigure}[b]{0.19\textwidth}
         \centering
         \includegraphics[width=1\textwidth]{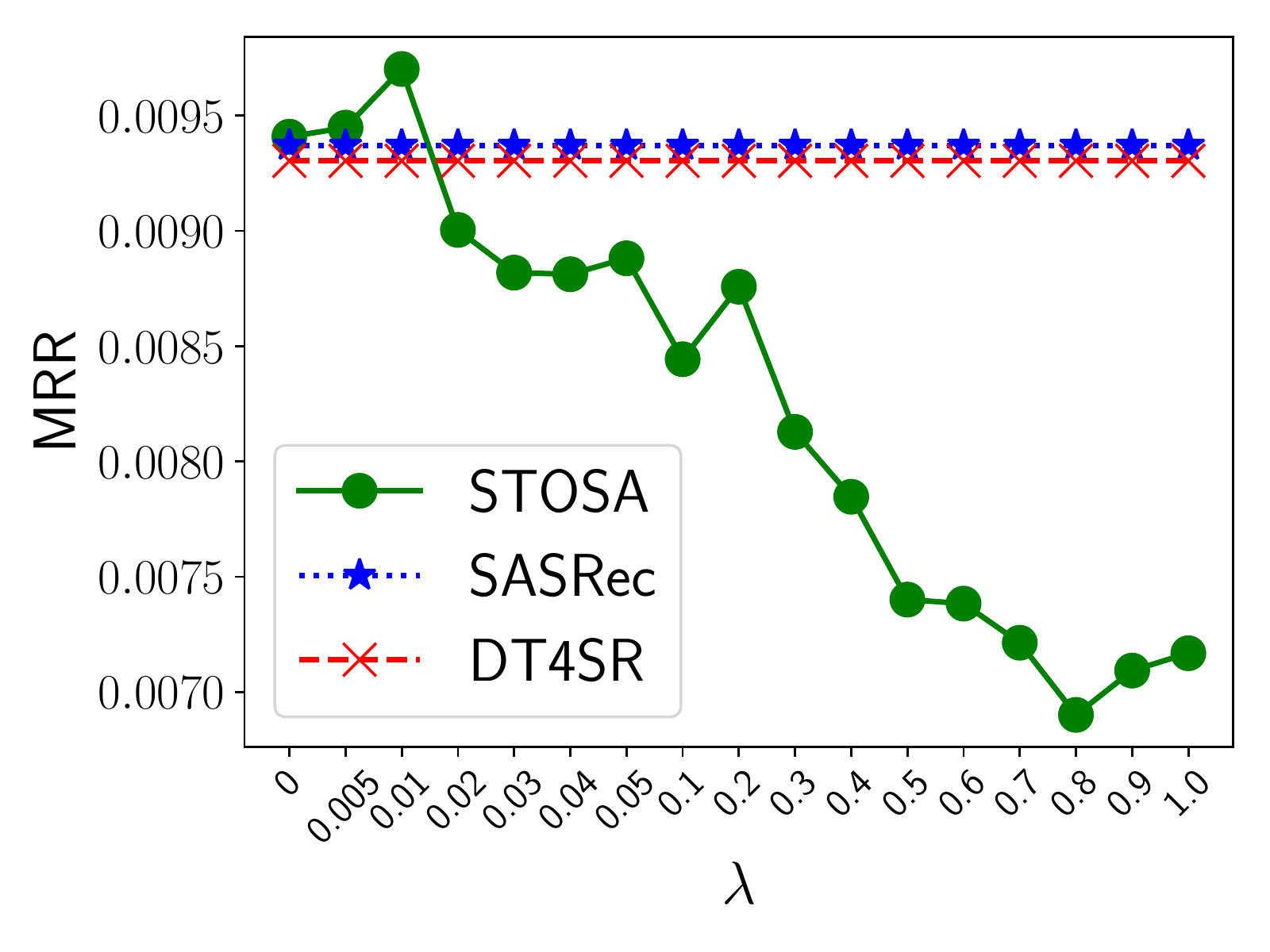}
         \caption{Home}
         \label{fig:mrr_toy}
     \end{subfigure}\hfill
     \begin{subfigure}[b]{0.19\textwidth}
         \centering
         \includegraphics[width=1\textwidth]{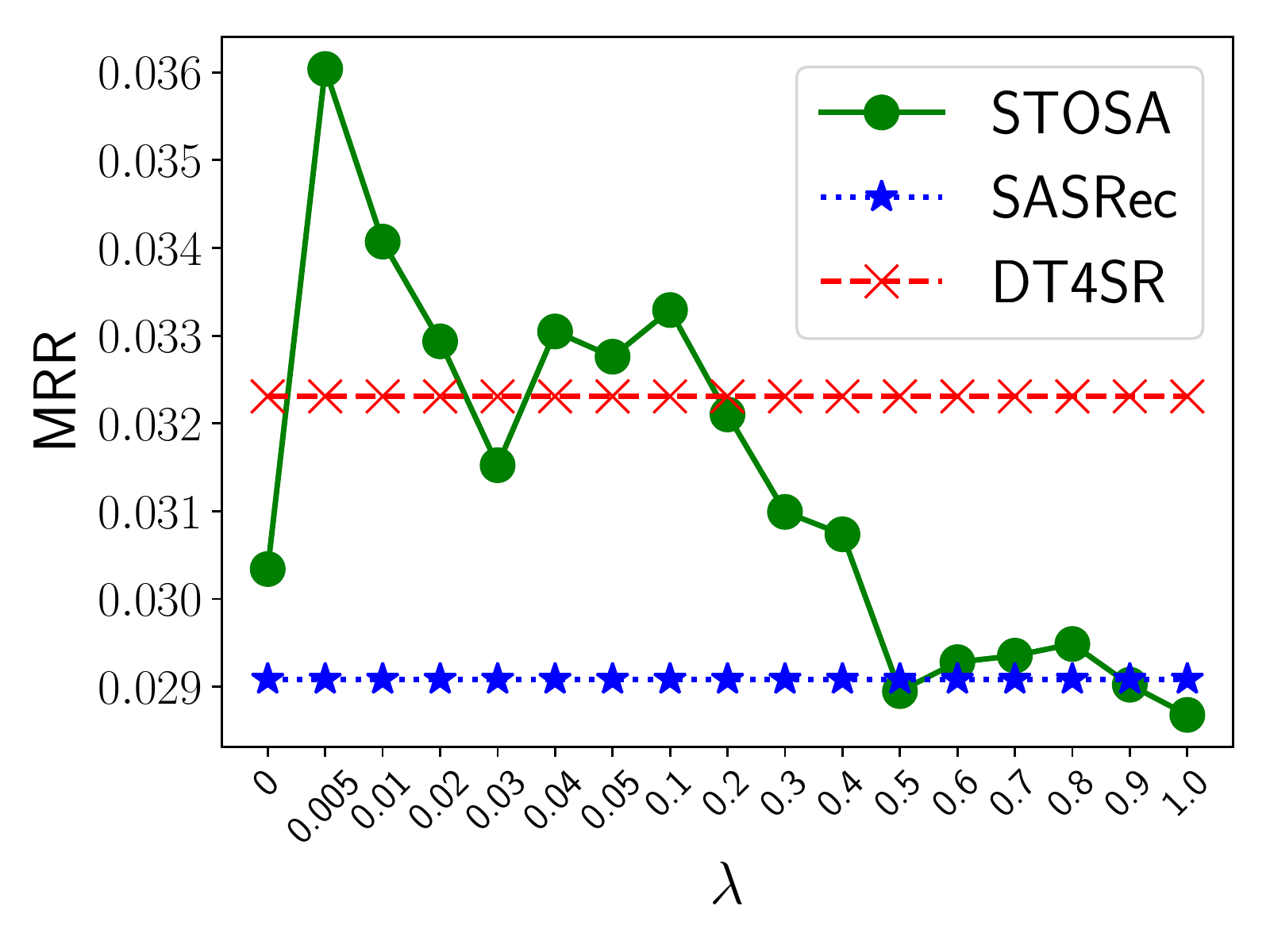}
         \caption{Beauty}
         \label{fig:mrr_baby}
     \end{subfigure}\hfill
     \begin{subfigure}[b]{0.19\textwidth}
         \centering
         \includegraphics[width=1\textwidth]{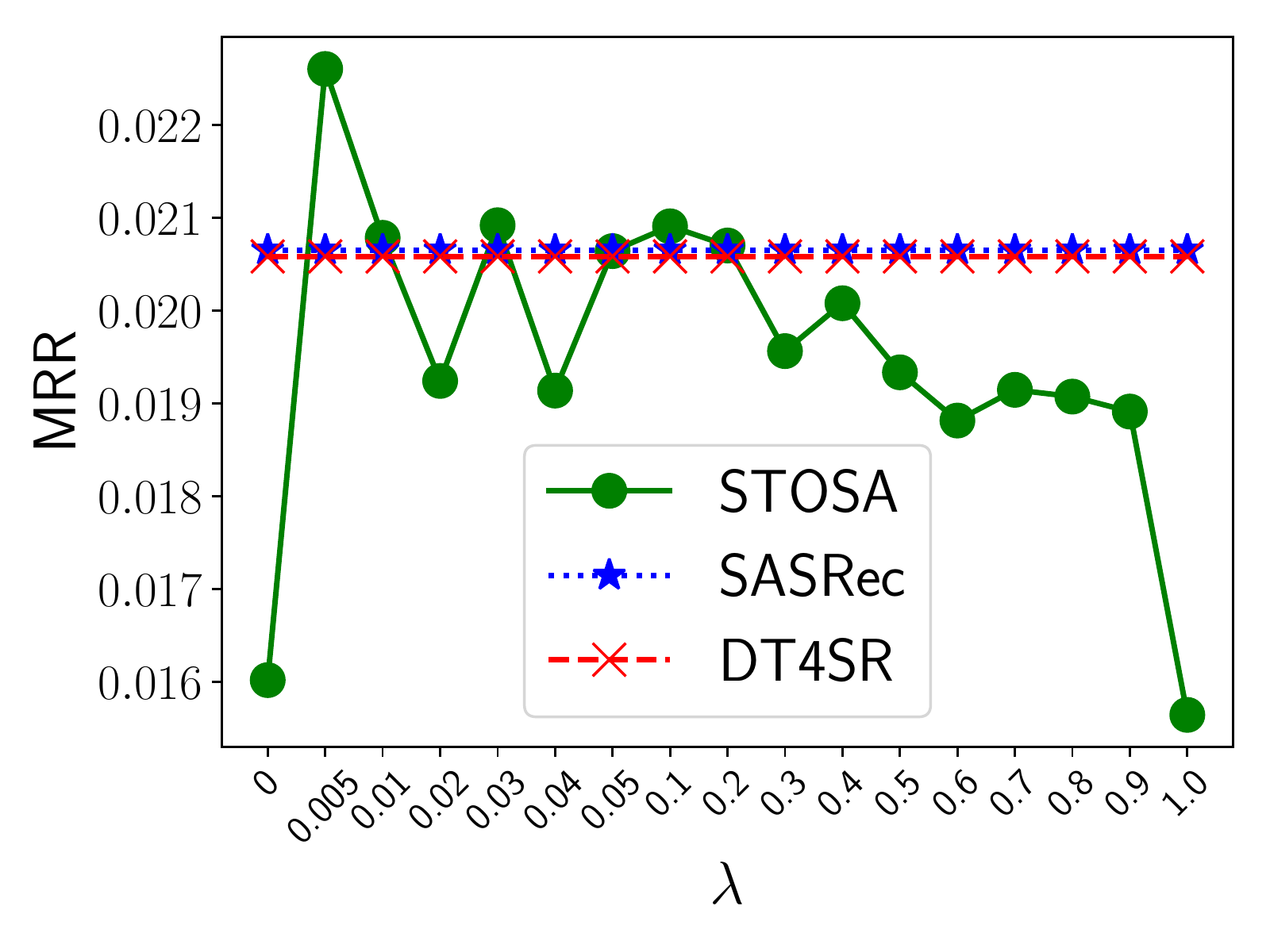}
         \caption{Tools}
         \label{fig:mrr_tools}
     \end{subfigure}\hfill
     \begin{subfigure}[b]{0.19\textwidth}
         \centering
         \includegraphics[width=1\textwidth]{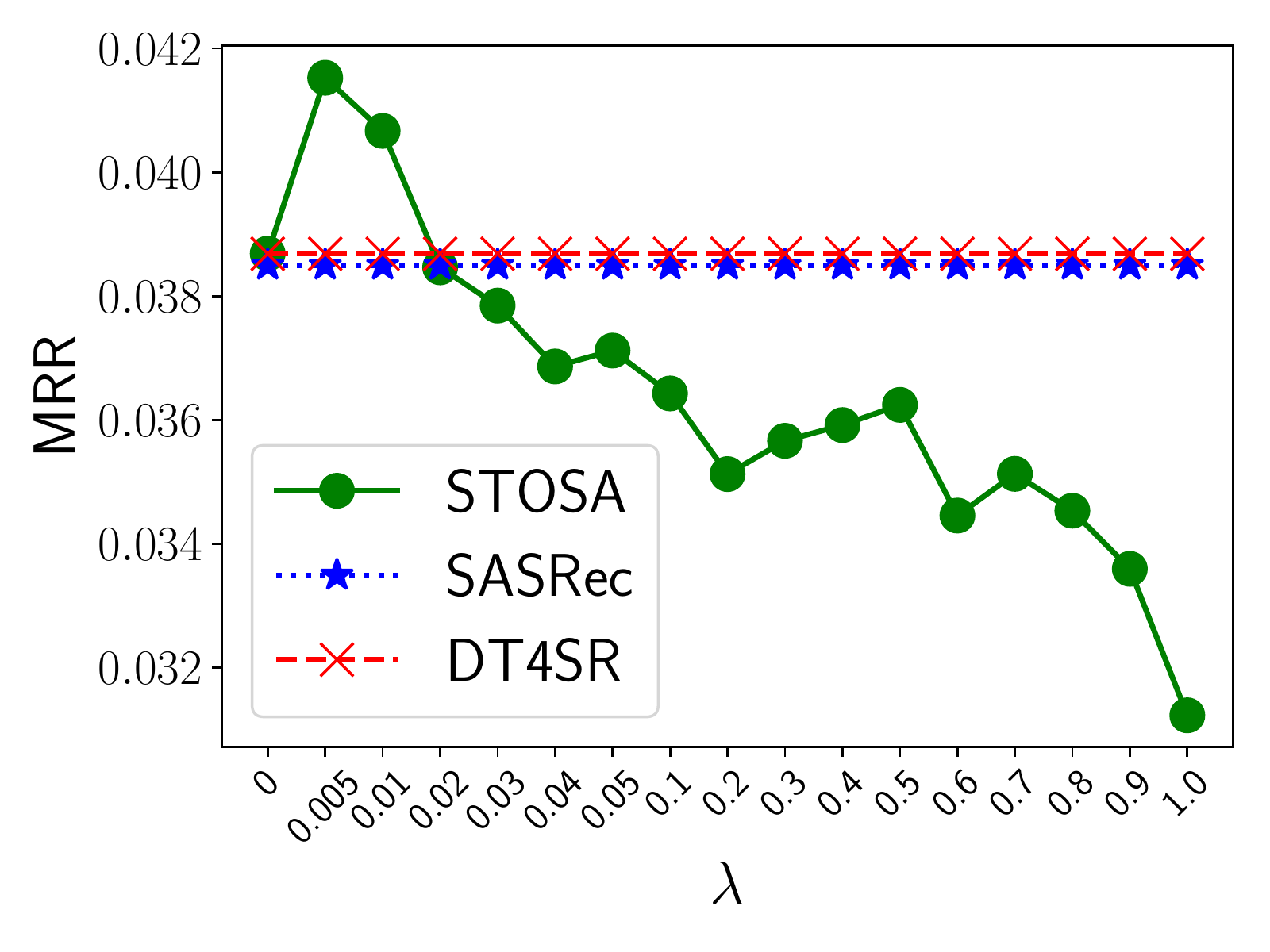}
         \caption{Toys}
         \label{fig:mrr_music}
     \end{subfigure}\hfill
     \begin{subfigure}[b]{0.19\textwidth}
         \centering
         \includegraphics[width=\textwidth]{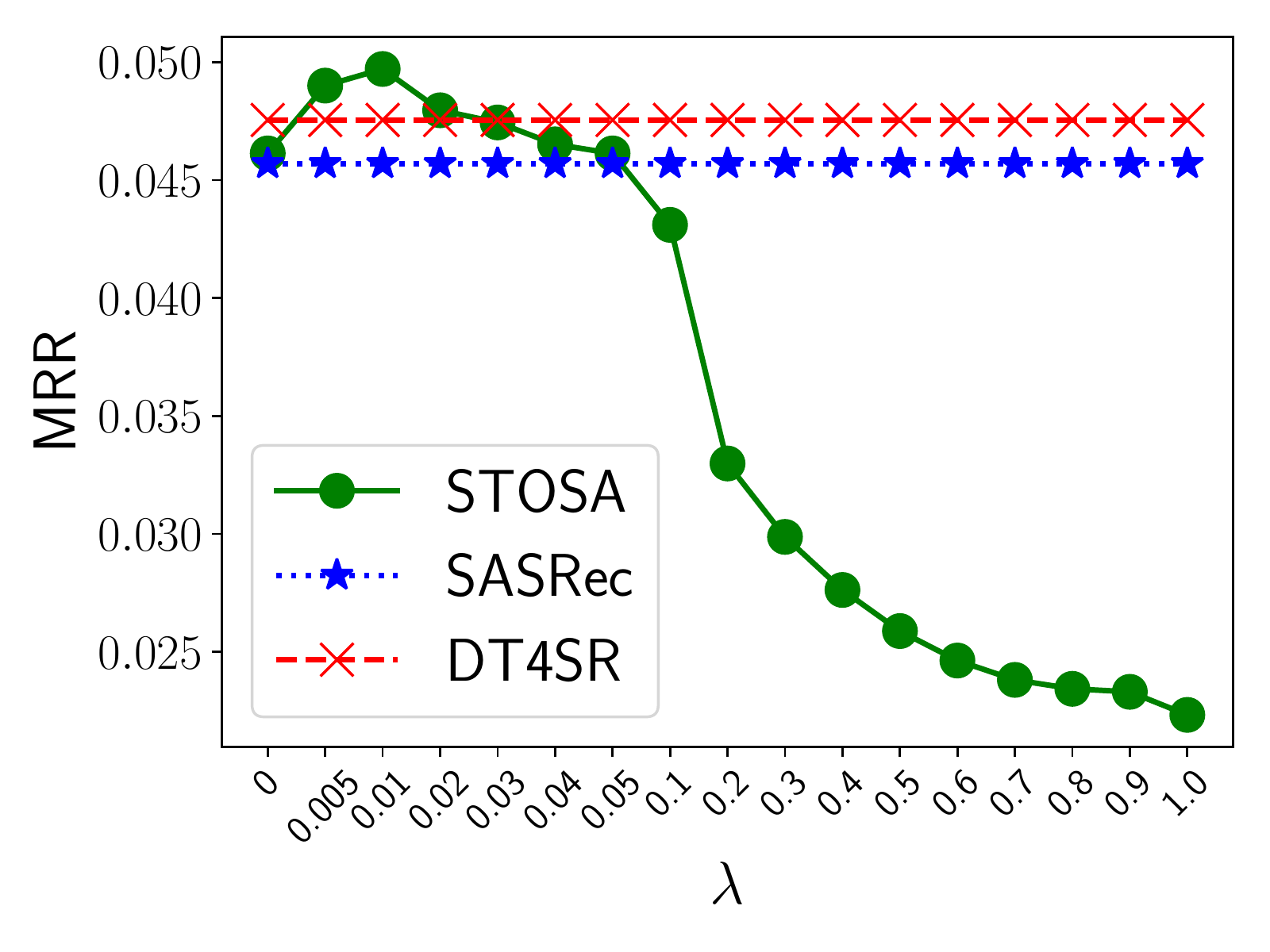}
         \caption{Office}
         \label{fig:mrr_office}
     \end{subfigure}
     \vspace{-3mm}
        \caption{MRR performances over various $\lambda$ on all datasets.}
        \label{fig:mrr_lambda}
\end{figure*}

\subsection{Parameters Sensitivity~(RQ2)}
In this section, we investigate the performance sensitivity of the weight $\lambda$ on the additional regularization $\ell_{pvn}$ across all datasets. Recall that the $\lambda$ in Eq.~(\ref{eq:loss_obj}) constraints the distance between the positive items and sampled negative items to be no less than the ground truth prediction distance. The trends are shown in Figure~\ref{fig:mrr_lambda}. 

With a proper selection of $\lambda$, \modelname can perform better than SASRec and DT4SR. We can observe that as the values of $\lambda$ become larger, the MRR performance first improves then drops. Another observation is that the values of $\lambda$ can significantly affect the performance. A properly selected $\lambda$ can dramatically improves the performance, indicating the necessity of the consideration of distances between positive items and sample negative items. However, $\lambda$ should not be too large as negative items are not strictly negative rather than sampled negative items, which is the potential reason why the performance drops when $\lambda$ increases. 

One special case is setting $\lambda=0$, which is also an ablation study of verifying the effectiveness of \modelname. When $\lambda=0$, \modelname still outperforms SASRec in most datasets, except the Tools dataset. This indicates the superiority of Wasserstein Self-Attention and the necessity of modeling uncertainty information and collaborative transitivity for the sequential recommendation. 
\subsection{Improvements Analysis~(RQ3)}
In this section, we analyze the sources of performance gains by comparing with SASRec on different groups of users and items. The analysis verifies the effectiveness of uncertainty information in user modeling and cold start items issue alleviation.
\begin{figure*}[!ht]
     \centering
     \begin{subfigure}[b]{0.19\textwidth}
         \centering
         \includegraphics[width=1\textwidth]{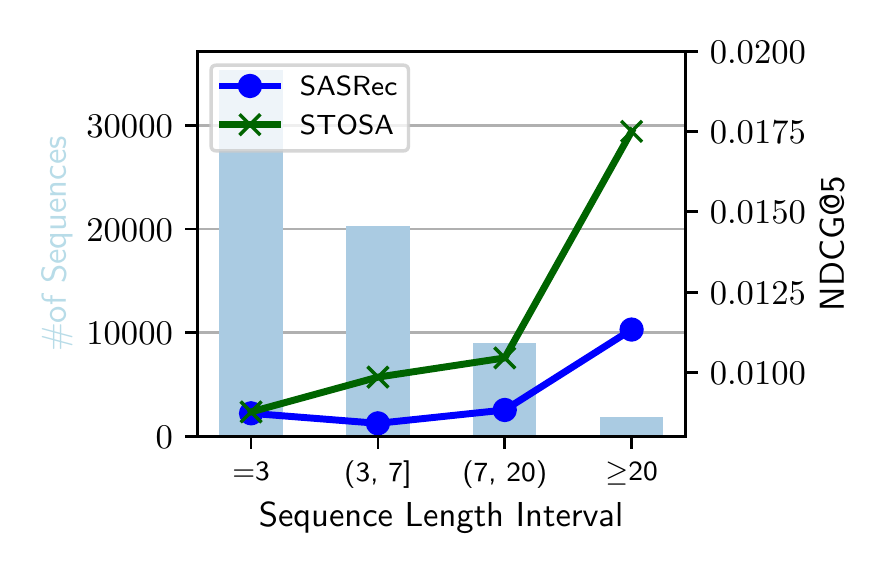}
         \caption{Home}
         \label{fig:mrr_home}
     \end{subfigure}\hfill
     \begin{subfigure}[b]{0.19\textwidth}
         \centering
         \includegraphics[width=1\textwidth]{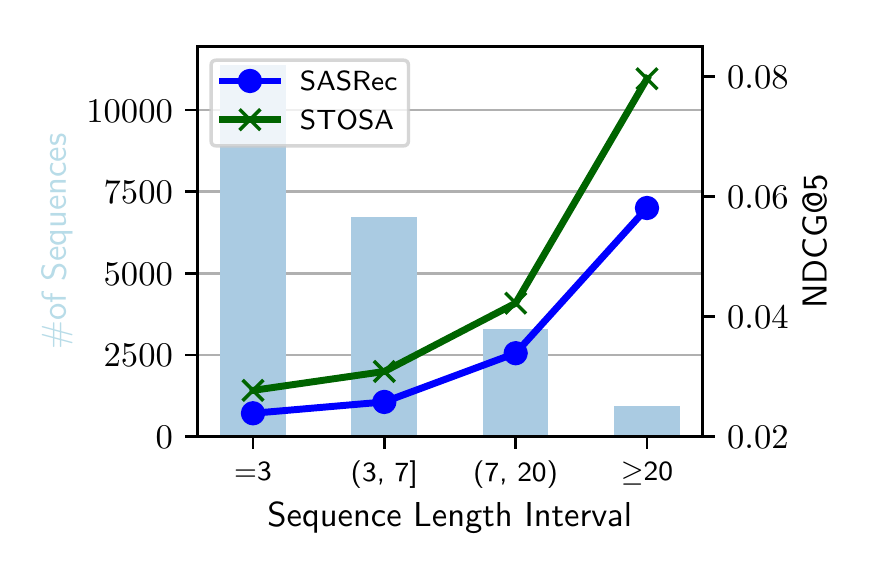}
         \caption{Beauty}
         \label{fig:mrr_beauty}
     \end{subfigure}\hfill
     \begin{subfigure}[b]{0.19\textwidth}
         \centering
         \includegraphics[width=1\textwidth]{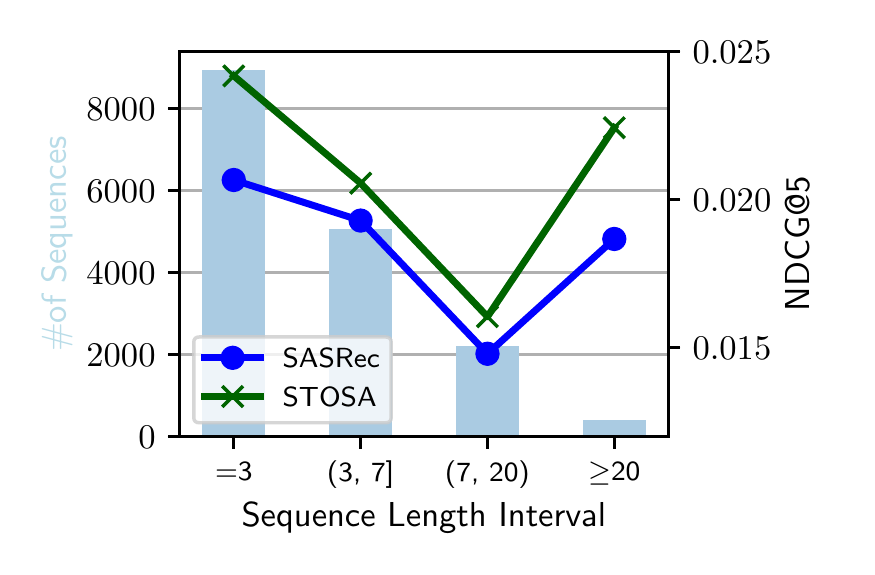}
         \caption{Tools}
         \label{fig:mrr_tools}
     \end{subfigure}\hfill
     \begin{subfigure}[b]{0.19\textwidth}
         \centering
         \includegraphics[width=1\textwidth]{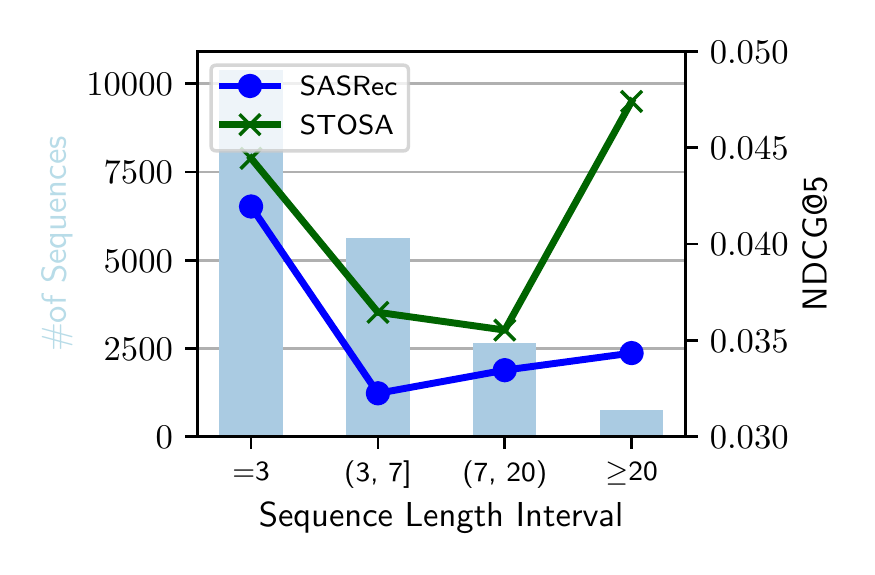}
         \caption{Toys}
         \label{fig:mrr_toys}
     \end{subfigure}\hfill
     \begin{subfigure}[b]{0.19\textwidth}
         \centering
         \includegraphics[width=\textwidth]{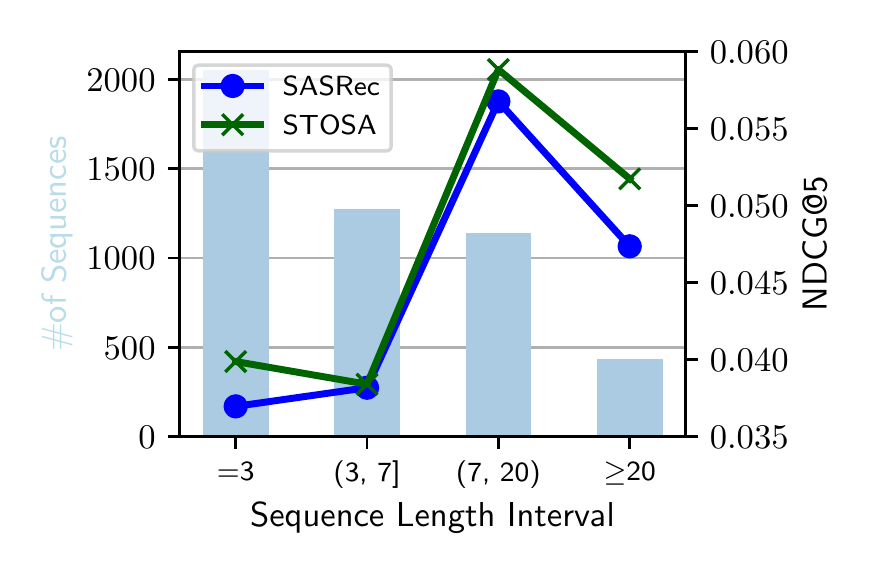}
         \caption{Office}
         \label{fig:mrr_office}
     \end{subfigure}
     \vspace{-3mm}
        \caption{NDCG@5 performances on different sequence lengths~(\textit{i.e., } number of training interactions of users) on all datasets.}
        \label{fig:ndcg5_seqlen}
\end{figure*}
\subsubsection{Performances w.r.t Sequence Lengths}
We separate users into groups based on their number of interactions in the training portion, which is also the training sequence lengths of users. We report the average NDCG@5  on each group of users. Figure~\ref{fig:ndcg5_seqlen} shows the sizes of each group of users and the corresponding NDCG@5 performances. The group with the shortest sequence length has the most users, and sizes decrease as sequence lengths become longer. 

From Figure~\ref{fig:ndcg5_seqlen}, \modelname achieves most significant improvements in users within the largest sequence length interval, compared with short sequences. The relative improvements on the longest sequence interval range from 9.70\% to 54.45\% across all datasets. The intuition behind these improvements is that users with more interactions are more likely to have diverse interests, indicating more uncertain behaviors. It demonstrates the effectiveness of stochastic representations in capturing uncertainty in user behaviors. We can also observe that \modelname can achieve comparative and better performances in most sequence length intervals, demonstrating the superiority of \modelname for the sequential recommendation.
\begin{figure*}[ht]
     \centering
     \begin{subfigure}[b]{0.19\textwidth}
         \centering
         \includegraphics[width=1\textwidth]{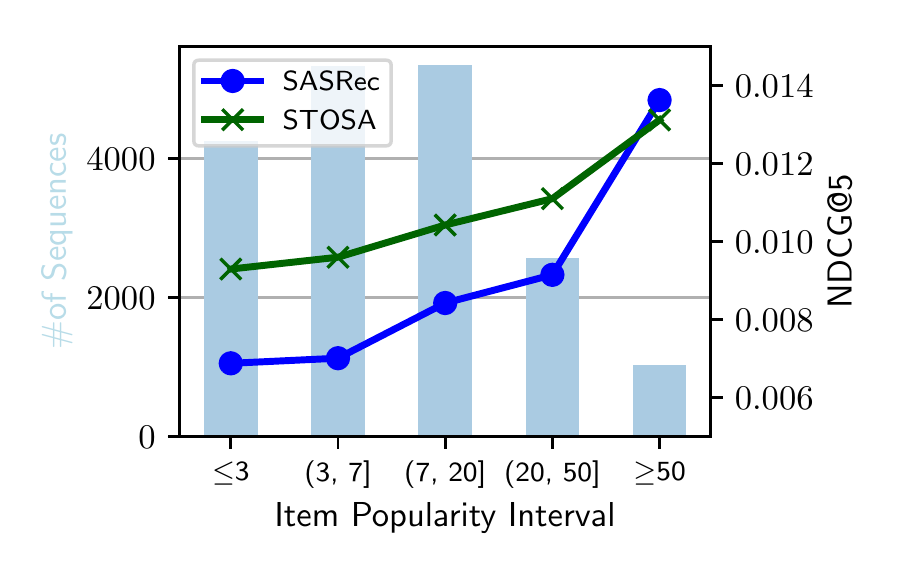}
         \caption{Home}
         \label{fig:mrr_home}
     \end{subfigure}\hfill
     \begin{subfigure}[b]{0.19\textwidth}
         \centering
         \includegraphics[width=1\textwidth]{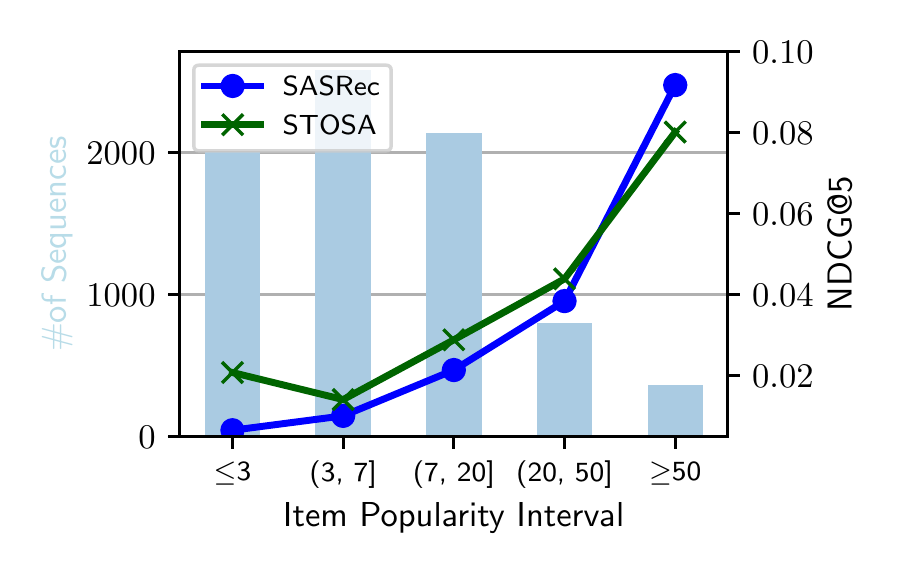}
         \caption{Beauty}
         \label{fig:mrr_beauty}
     \end{subfigure}\hfill
     \begin{subfigure}[b]{0.19\textwidth}
         \centering
         \includegraphics[width=1\textwidth]{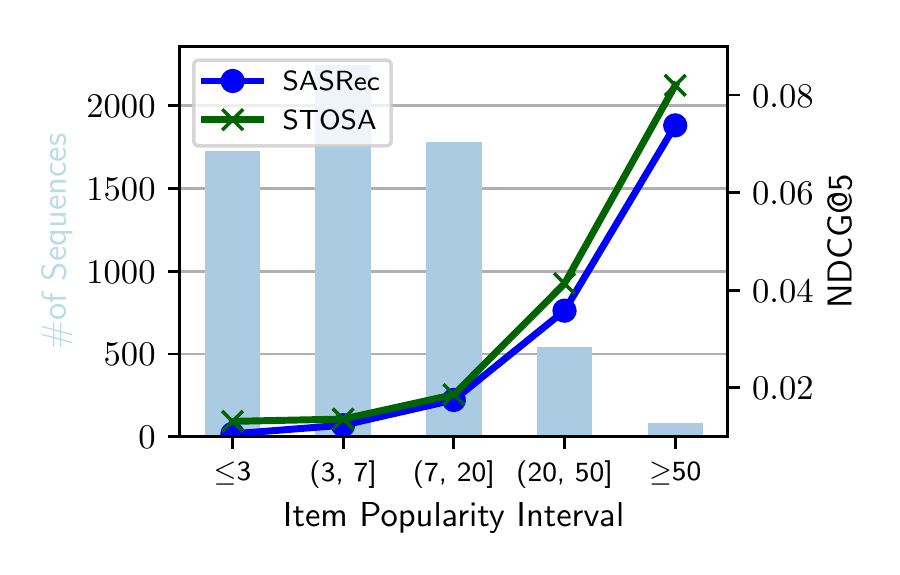}
         \caption{Tools}
         \label{fig:mrr_tools}
     \end{subfigure}\hfill
     \begin{subfigure}[b]{0.19\textwidth}
         \centering
         \includegraphics[width=1\textwidth]{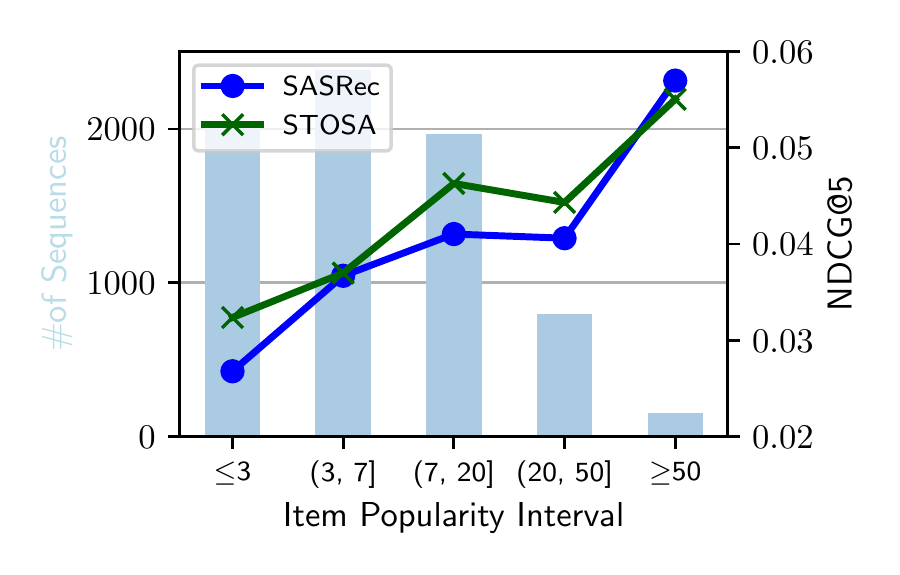}
         \caption{Toys}
         \label{fig:mrr_toys}
     \end{subfigure}\hfill
     \begin{subfigure}[b]{0.19\textwidth}
         \centering
         \includegraphics[width=\textwidth]{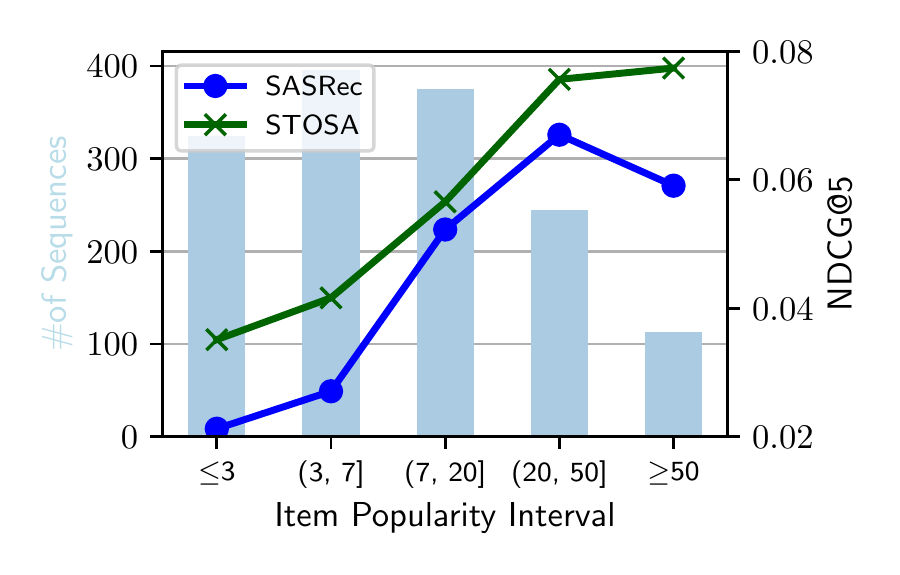}
         \caption{Office}
         \label{fig:mrr_office}
     \end{subfigure}
     \vspace{-3mm}
        \caption{NDCG@5 performances on different item popularity~(\textit{i.e., } number of training interactions of items) on all datasets.}
        \label{fig:ndcg5_itempop}
\end{figure*}
\subsubsection{Performances w.r.t Item Popularity}
We investigate the performances on different groups of items based on the popularity to demonstrate that the collaborative transitivity and the proposed regularization $\ell_{pvn}$ both help alleviate the cold-start items issue. We report the sizes and average NDCG@5 on each group of items, and each group is separated based on popularity. The distribution of sizes is similar to the one of users, where most items are unpopular. 

The performances comparison on all datasets is shown in Figure~\ref{fig:ndcg5_itempop}. For all datasets, the best improvements are from items with interactions no more than 3~(\textit{i.e.,} cold start items). This observation supports the effectiveness of Wasserstein Self-Attention in capturing collaborative transitivity and the additional regularization $\ell_{pvn}$ in generalizing the latent item transitions discovery. However, the performance becomes worse for popular items in Beauty and Toys datasets. We believe the reason might be the noisy neighbors of popular items in stochastic representations, which introduces potentially larger space for collaborative neighborhoods discovery.

\subsection{Qualitative Analysis}
In this section, we qualitatively visualize the attention weights and some examples of similar items retrieval. Specifically, we conduct a case study on a specific user with a long sequence length while her test item is a cold start item, which helps identify the significant difference between \modelname and SASRec. We also analyze the proposed \modelname by comparing the prediction lists with SASRec and \modelname, which can be found in Appendices.

\subsubsection{Wasserstein Self-Attentions Visualization~(RQ4)}
Figure~\ref{fig:att_weights} in the Appendices illustrates the heat maps of self-attention weights on the last 20 positions,  learned by SASRec and \modelname, respectively. Recall that one of the critical differences between SASRec and \modelname is the calculation of attention weights, where SASRec adopts dot-product, and \modelname uses negative 2-Wasserstein distance. 

We can observe some commonalities and differences between two attention weights heat maps. The attention weights of \modelname and SASRec are shown in Figure~\ref{fig:att_weights_stosa} and Figure~\ref{fig:att_weights_sasrec}, respectively. Both attention weights give larger weights to more recent behaviors, where the values in the bottom right corner are significant. However, \modelname has a more uniform attention weights distribution than SASRec as SASRec only highlights a small set of items in the sequence. The reason behind this difference is potentially the consideration of collaborative transitivity, which connects co-occurred items more tightly and introduces more collaborative neighbors in item-item transitions modeling.

\subsubsection{Item Embeddings Visualization Comparison~(RQ5)}
We use T-SNE~\cite{van2008visualizing} to visualize the latent spaces of items learned by SASRec and \modelname, respectively, as shown in Figure~\ref{fig:item_emb_visual} in the Appendices. We color items based on the items' popularity. We have the following observations: (1). the distributions of popular items~(items with popularity more than 7) are significantly different between SASRec and \modelname, and (2). the distributions of cold items~(popularity $<= 3$) are mostly uniform in both SASRec and \modelname. In SASRec, popular items are mostly located far away from the center and not closely connected to each other. \modelname learns a significantly different distribution of popular items, forcing them to locate close to the center and form a denser connected group. Moreover, due to the limited data for cold items, most cold items are uniformly distributed.

This difference is attributed to the collaborative transitivity learned in \modelname, which SASRec ignores and fails to generalize collaborative signals to cold items. The collaborative transitivity helps cold items to retrieve more collaborative neighbors as popular items can be related to triangle inequality. It again demonstrates the necessity and superiority of collaborative transitivity signals for the sequential recommendation.

\section{Conclusion}
This work proposes a novel stochastic self-attention sequential model \modelname for modeling dynamic uncertainty and capturing collaborative transitivity. We also introduce a novel regularization to BPR loss, guaranteeing a large distance between the positive item and negative sampled items. Extensive results and qualitative analysis on five real-world datasets demonstrate the effectiveness of \modelname and also well support the superiority of \modelname in alleviating cold start item recommendation issues as well as the necessity of collaborative transitivity for sequential recommendation. 

\section{Acknowledgments}
This work was supported in part by NSF under grants III-1763325, III-1909323,  III-2106758, and SaTC-1930941.
Hao Peng was supported by S\&T Program of Hebei through grant 21340301D.
For any correspondence, please refer to Ziwei Fan and Hao Peng.

\bibliographystyle{ACM-Reference-Format}
\balance
\bibliography{ref}

\newpage
\appendix
\appendixpage
\section{Complexity Analysis}
We analyze the space and time complexity of \modelname and demonstrate that \modelname has similar asymptotic space and time complexity as SASRec~\cite{kang2018self}. Note that even if stochastic embeddings are comprised of mean and covariance embeddings, we still use the same latent size as SASRec by equally separating dimensions to mean and covariance. For example, if we use $d=128$ in SASRec, we use $d_{\mu}=d/2=64$ and $d_{\Sigma}=d/2=64$ for fair comparisons.

\subsection{Space Complexity:} The learnable parameters in \modelname are from the stochastic embeddings and parameters in the Wasserstein self-attention layers, feed-forward networks and layer normalization. The overall number of parameters is $O(2|\mathcal{V}|\frac{d}{2} + 2n\frac{d}{2}+2(\frac{d}{2})^2) = O(|\mathcal{V}|d + nd + d^2/2)$, which is slightly smaller than the complexity of SASRec, which is $O(|\mathcal{V}|d+nd+d^2)$~\cite{kang2018self}.

\subsection{Time Complexity:} The computational complexity of \modelname is dominated by the Wasserstein self-attention layer and the feed-forward networks. The Wasserstein self-attention defined in Eq.~(\ref{eq:wass_dist_att}) can be converted to using batch matrix multiplications. The second term in Eq.~(\ref{eq:wass_dist_att}) can be transformed as a calculation of Euclidean norm as follows
\begin{equation}
    \text{trace}\left(\Sigma_{s_t}+\Sigma_{s_k}-2(\Sigma_{s_k}^{1/2}\Sigma_{s_t}\Sigma_{s_k}^{1/2})^{1/2}\right) = ||\Sigma_{s_t}^{1/2}-\Sigma_{s_k}^{1/2}||^2_{F},
\end{equation}
where $||\cdot||_F^2$ is Frobenius norm that can be calculated by matrix multiplications. Also, as $\Sigma_{s_t}$ and $\Sigma_{s_k}$ are both diagonal matrices, we can further reduce the computational complexity to $O(\frac{nd}{2}+\frac{n^2d}{2}+2n^2)$. And the Euclidean norm of mean embeddings part in Eq.~(\ref{eq:wass_dist_att}) can also be calculated by matrix multiplications with the same time complexity. Therefore, the overall Wasserstein self-attention time complexity is $O(nd+n^2d+4n^2)$. By also considering the feed-forward networks, we obtain the final asymptotic computational complexity as $O(nd+n^2d+4n^2+\frac{nd^2}{2})$.
The computation complexity of traditional self-attention~\cite{kang2018self} is $O(n^2d+nd^2)$.
Note that both complexities are typically dominated by the $O(n^2d)$ term as $d$ is typically much larger than 4, It indicates that \modelname has asymptotic similarly time complexity with SASRec.

\begin{table}[H]
\centering
\caption{Datasets Statistics}
\label{tab:data_stat}
\resizebox{0.48\textwidth}{!}{%
\begin{tabular}{@{}crrrrc@{}}
\toprule
Dataset & \multicolumn{1}{c}{\#users} & \multicolumn{1}{c}{\#items} & \multicolumn{1}{c}{\#interactions} & \multicolumn{1}{c}{density} & \begin{tabular}[c]{@{}c@{}}avg. \\ interactions \\ per user\end{tabular} \\ \midrule
Home & 66,519 & 28,237 & 551,682 & 0.03\% & 8.3 \\
Beauty & 22,363 & 12,101 & 198,502 & 0.05\% & 8.3 \\
Toys & 19,412 & 11,924 & 167,597 & 0.07\% & 8.6 \\
Tools & 16,638 & 10,217 & 134,476 & 0.08\% & 8.1 \\
Office & 4,905 & 2,420 & 53,258 & 0.44\% & 10.8 \\
\bottomrule
\end{tabular}%
}
\end{table}
\section{Datasets and Preprocessing}
Details of datasets statistics\footnote{\url{https://jmcauley.ucsd.edu/data/amazon/}} are presented in Table~\ref{tab:data_stat}.

\section{Implementation Details and Baselines Grid Search}
We implement \modelname with Pytorch in a Nvidia 3090 GPU with 64GB system memory. We grid search all parameters and report the test performance based on the best validation results. For all baselines, we search the embedding dimension in $\{64, 128\}$. As the proposed model has both mean and covariance embeddings, we only search for $\{32, 64\}$ for \modelname for the fair comparison. We also search max sequence length from $\{50, 100\}$.  We tune the learning rate in $\{10^{-3},10^{-4}\}$, search the L2 regularization weight from $\{10^{-1}, 10^{-2}, 10^{-3}\}$, dropout rate from $\{0.3, 0.5, 0.7\}$. For sequential methods, we search number of layers from $\{1,2,3\}$, and number of heads in $\{1,2,4\}$. We adopt the early stopping strategy that model optimization stops when the validation MRR does not increase for 50 epochs. The followings are the model specific hyper-parameters search ranges of baselines:
The third group consists of sequential recommendation methods:
\begin{itemize}[leftmargin=*]
    \item \textbf{BPR\footnote{\url{https://github.com/xiangwang1223/neural_graph_collaborative_filtering}}:} BPR is the most classical collaborative filtering method for personalized ranking with implicit feedbacks. We search the learning rate in $\{10^{-3},10^{-4}\}$ and L2 regularization weight from $\{10^{-1}, 10^{-2}, 10^{-3}\}$.
    \item \textbf{LightGCN\footnote{\url{https://github.com/kuandeng/LightGCN}}:} LightGCN is the state-of-the-art static recommendation method, which considers high-order collaborative signals in user-item graph. We search number of layers from $\{1,2,3\}$, and node dropout from $\{0.1, 0.3, 0.5, 0.7\}$.
    \item \textbf{CML\footnote{\url{https://github.com/changun/CollMetric}}:} One of the earliest works that adopt distance metrics to measure the affinity between users and items for recommendation. We search covariance loss weight from $\{0.1, 0.3\}$, and margin from $\{3.0, 4.0, 5.0\}$.
    \item \textbf{SML\footnote{\url{https://github.com/MingmingLie/SML}}:} This method is the state-of-the-art metric learning recommendation method. It extends CML to additionally consider item-centric distances. We search the $\gamma$ from $\{5, 10, 20\}$, $\lambda$ from $\{0.01, 0.1, 1.0\}$, and margin $\{0.1, 0.3, 0.5\}$.
    \item \textbf{TransRec\footnote{\url{https://github.com/YifanZhou95/Translation-based-Recommendation}}:} A metric learning-based sequential recommendation that proposes translation vectors to encode the item transition relationships. We search the $\lambda$ from $\{0.001, 0.01, 0.1\}$.
    \item \textbf{Caser\footnote{\url{https://github.com/graytowne/caser_pytorch}}:} A CNN-based sequential recommendation method that views the sequence embedding matrix as an image and applies convolution operators to it. We search the length $L$ from $\{5, 10\}$, and $T$ from $\{1, 3, 5\}$.
    \item \textbf{SASRec\footnote{\url{https://github.com/RUCAIBox/CIKM2020-S3Rec}}:} The state-of-the-art sequential method that depends on the Transformer architecture. We search the dropout rate from $\{0.3, 0.5, 0.7\}$. 
    \item \textbf{DT4SR\footnote{\url{https://github.com/DyGRec/DT4SR}}:} A metric learning-base sequential method that models items as distributions and proposes mean and covariance Transformers. We search the dropout rate from $\{0.3, 0.5, 0.7\}$.
    \item \textbf{BERT4Rec\footnote{\url{https://github.com/FeiSun/BERT4Rec}}:} This method extends SASRec to model bidirectional item transitions with standard Cloze objective. We search the mask probability from the range of $\{0.1, 0.2, 0.3, 0.5, 0.7\}$.
\end{itemize}

\section{Qualitative Anlaysis}
\subsection{Predictions Comparison}
\begin{table}[!h]
\centering
\caption{Prediction lists comparison of user A278LEQK1TEPVB. The ground truth item is in red and each item is associated with its ID and the popularity. The names of items are described in the following second paragraph.}
\label{tab:pred_list_diff}
\resizebox{0.4\textwidth}{!}{%
\begin{tabular}{@{}ccc@{}}
\toprule
Model & SASRec & \modelname \\ \midrule
{Rank-1} & (Wraparound Labels, 12) & {\color[HTML]{FE0000} (WF-7620 Printer, 3)} \\
{Rank-2} & (WF-3640 Printer, 3) & (Wraparound Labels, 12) \\
{Rank-3} & (Locker, 3) & (WF-3640 Printer, 3) \\
{Rank-4} & (Binders, 1) & (Speaker Phone, 8) \\
{Rank-5} & {\color[HTML]{FE0000} (WF-7620 Printer, 3)} & (Ring Binder, 10) \\ \bottomrule
\end{tabular}%
}
\end{table}
We show the differences of top-5 predicted ranking lists for the user A278LEQK1TEPVB between SASRec and \modelname in Table.\ref{tab:pred_list_diff}. The user's last five interacted items are $\mathcal{S}[-5:]=[$`Ink Refillable', `Write 'N Wipe', `Magnetic Whiteboard', `Pencil Cup Holder', `WF-4640 Inkjet Printer' $]$.
Moreover, \modelname can prioritize more relevant items even when items are cold start. The last interacted item of the user A278LEQK1TEPVB is a Inkjet Printer. The rank-1 item of \modelname WF-7620 Printer is a better version of Inkjet Printer while the rank-1 item of SASRec is a Wraparound Labels, which does not match with the user's real interests. The reason might be that SASRec prefers more popular items and Wraparound Labels has more interactions than WF-7620 Printer. We can conclude that stochastic representations in \modelname can introduce different neighbors by utilizing collaborative transitivity information and also upvote relevant but cold start items in recommendation list.


\subsection{Wasserstein Self-Attentions Visualization~(RQ4)}
\begin{figure}[ht]
     \centering
     \begin{subfigure}[b]{0.235\textwidth}
         \centering
         \includegraphics[width=1\textwidth]{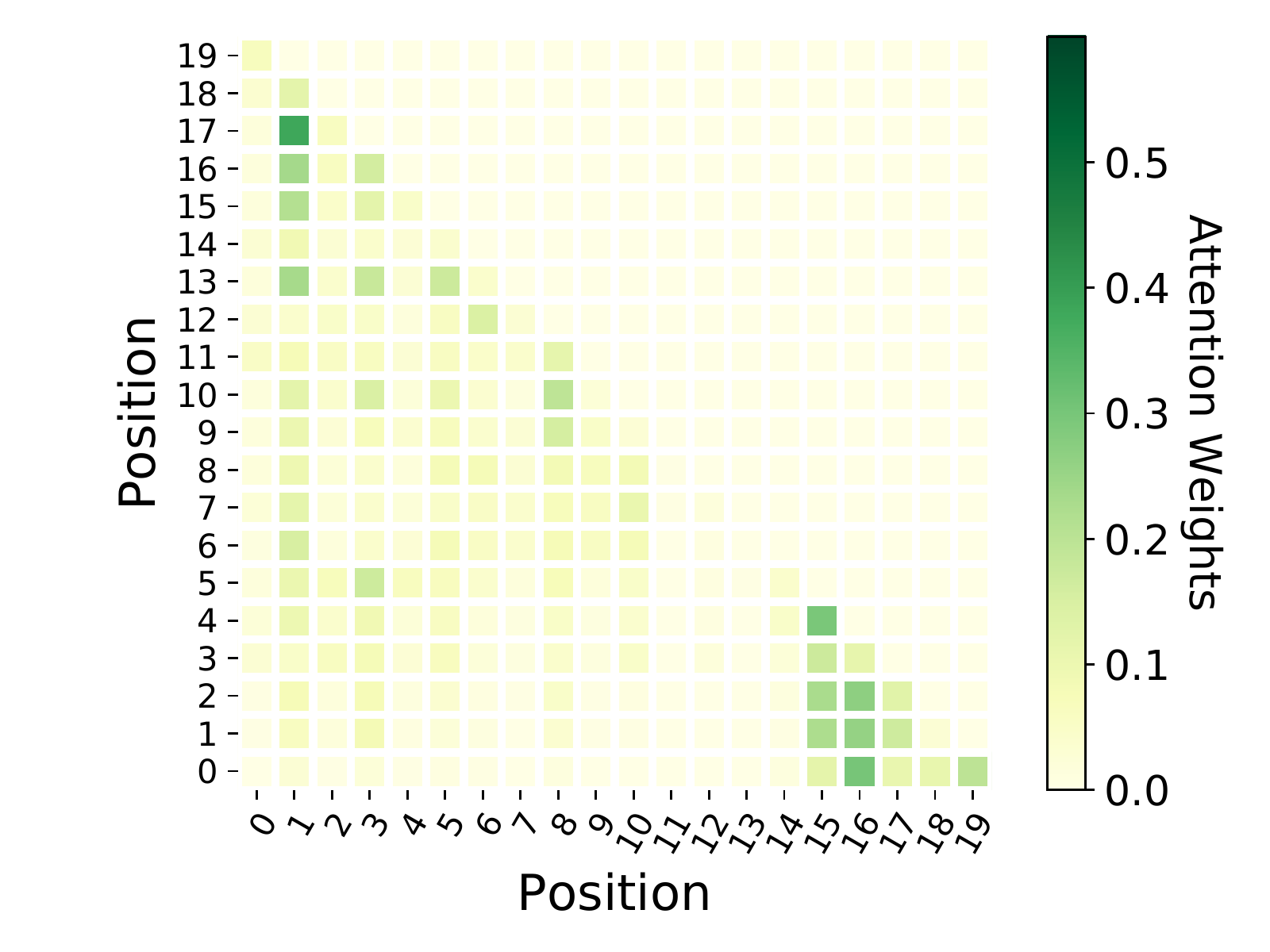}
         \caption{\modelname}
         \label{fig:att_weights_stosa}
     \end{subfigure}\hfill
     \begin{subfigure}[b]{0.235\textwidth}
         \centering
         \includegraphics[width=1\textwidth]{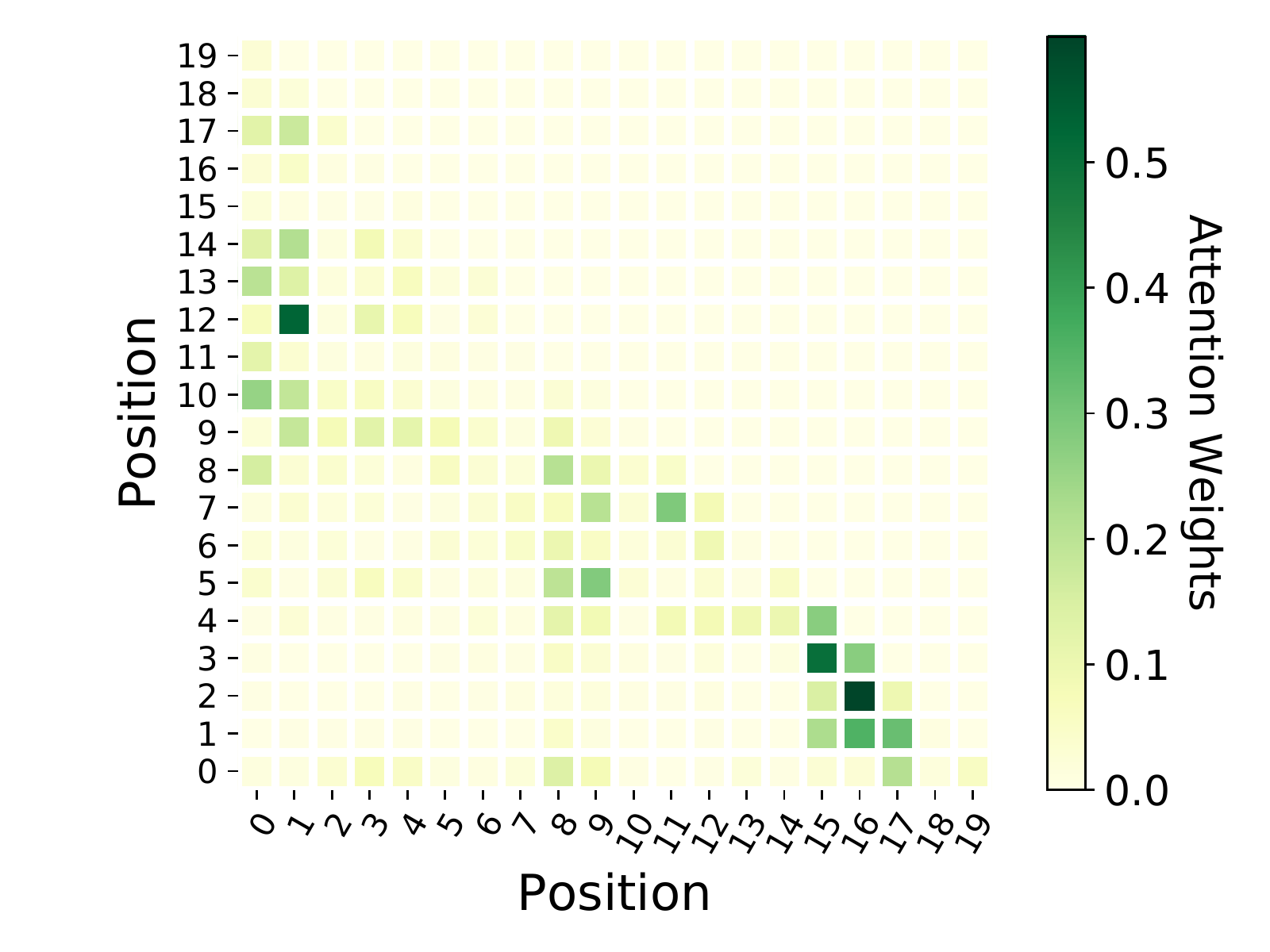}
         \caption{SASRec}
         \label{fig:att_weights_sasrec}
     \end{subfigure}
        \caption{Attention Weights Visualizations of \modelname and SASRec on the user A278LEQK1TEPVB in Office dataset.}
        \label{fig:att_weights}
\end{figure}

\subsection{Item Embeddings Visualization Comparison~(RQ5)}
\begin{figure}[ht]
     \centering
     \begin{subfigure}[b]{0.235\textwidth}
         \centering
         \includegraphics[width=1\textwidth]{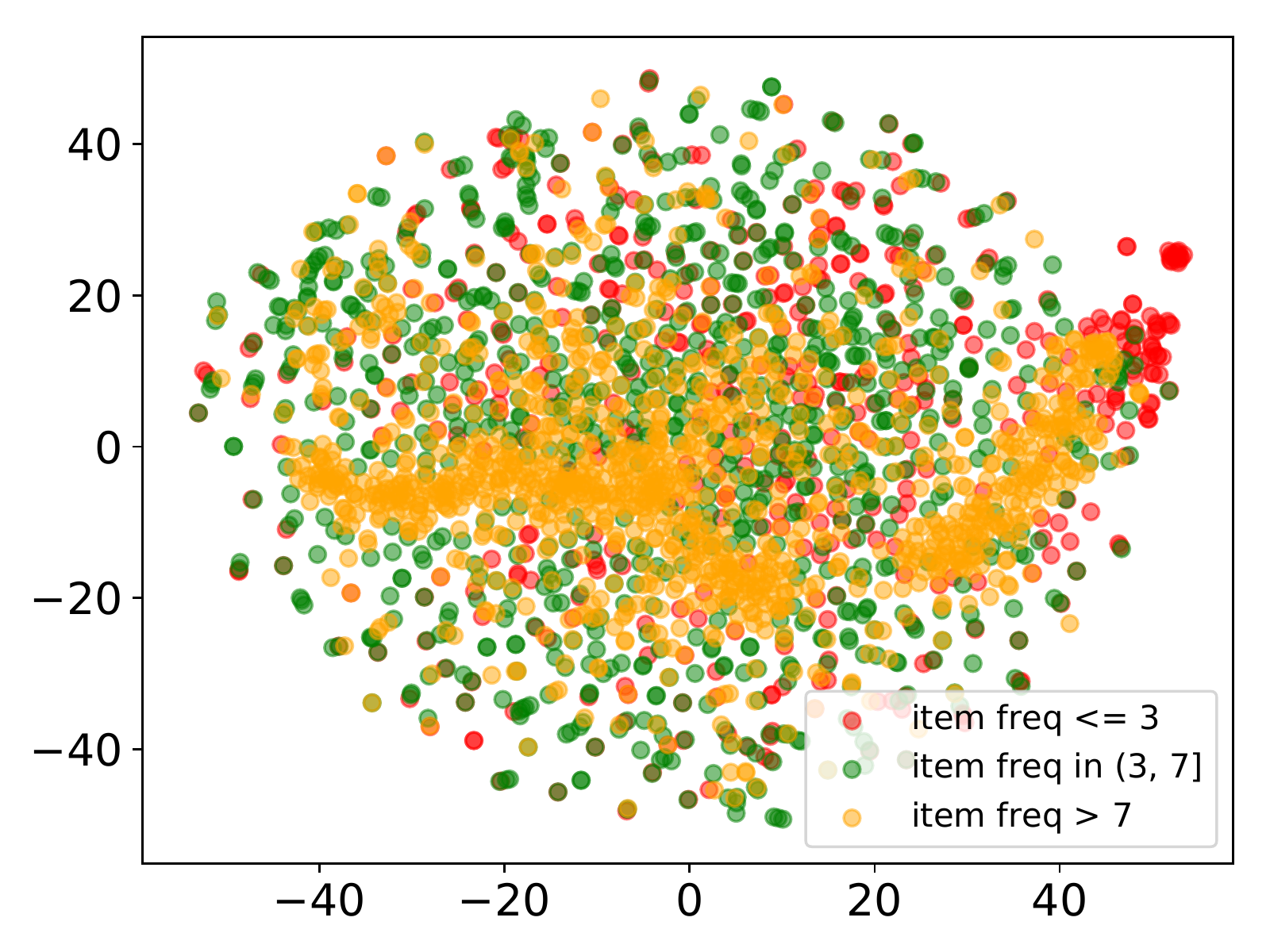}
         \caption{\modelname}
         \label{fig:item_emb_visual_stosa}
     \end{subfigure}\hfill
     \begin{subfigure}[b]{0.235\textwidth}
         \centering
         \includegraphics[width=1\textwidth]{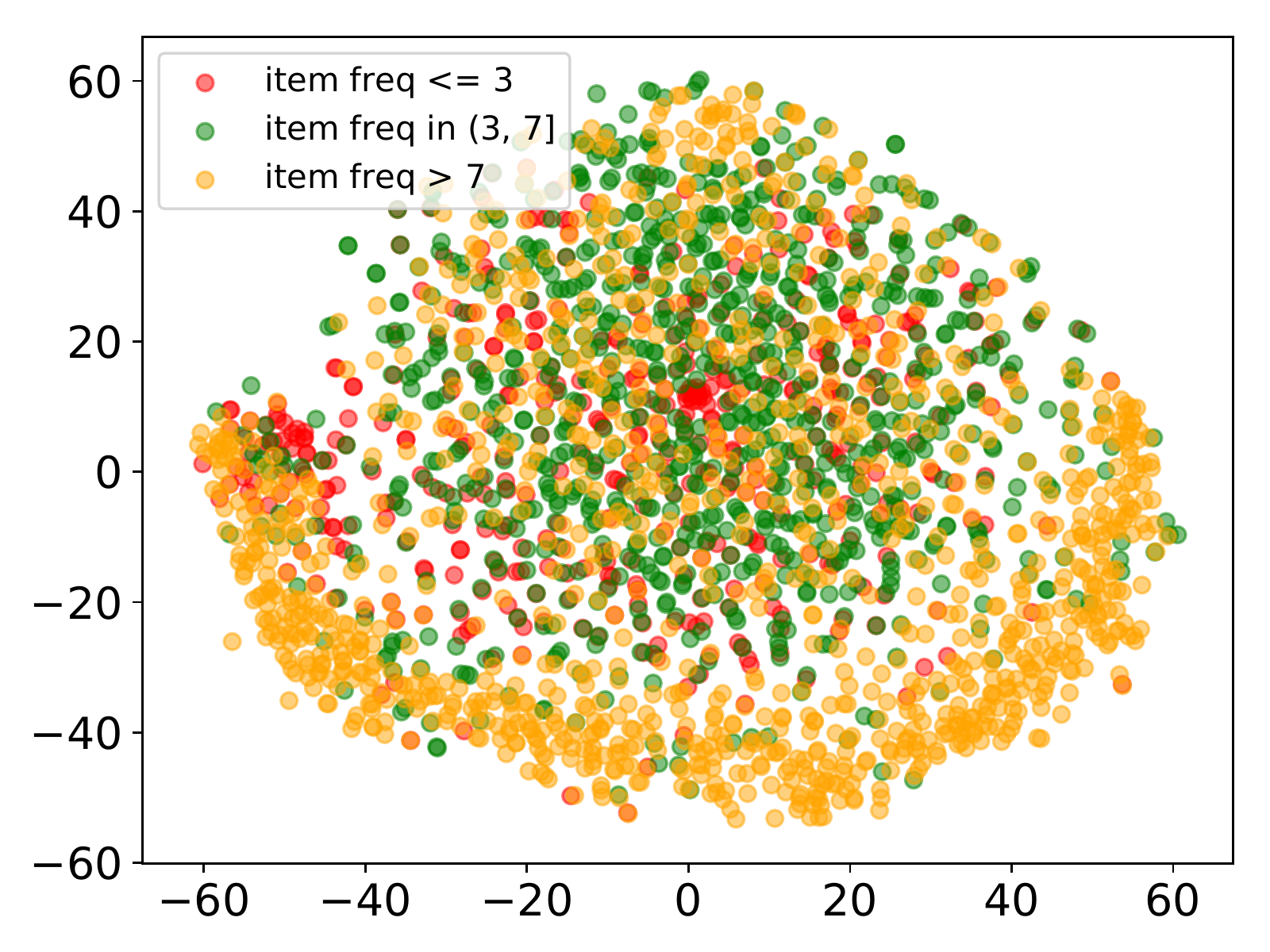}
         \caption{SASRec}
         \label{fig:item_emb_visual_sasrec}
     \end{subfigure}
        \caption{T-SNE Visualizations of item embeddings in Office dataset learned from \modelname and SASRec. The figures are best viewed in color.}
        \label{fig:item_emb_visual}
\end{figure}

\end{document}